\documentclass[journal,final,twoside]{IEEEtran}

\usepackage[noadjust]{cite}
\usepackage[dvipsnames]{xcolor}
\usepackage[normalem]{ulem}
\usepackage{cuted}
\usepackage[hidelinks]{hyperref}
\usepackage{dblfloatfix} % To enable figures at the bottom of page
\ifCLASSINFOpdf
  \usepackage{graphicx}
  \graphicspath{{figures/}{../jpeg/}}
  \DeclareGraphicsExtensions{.eps,.pdf,.jpeg,.png}
\else
\fi

\ifCLASSOPTIONcompsoc
    \usepackage[caption=false, font=normalsize, labelfont=sf, textfont=sf]{subfig}
\else
    \usepackage[caption=false, font=footnotesize]{subfig}
\fi

\usepackage{amsmath,amssymb,bm}
\usepackage{xfrac}
\usepackage{upgreek}
\usepackage{siunitx}
\DeclareSIUnit{\belmilliwatt}{Bm}
\DeclareSIUnit{\belwatt}{BW}
\DeclareSIUnit{\dBm}{\deci\belmilliwatt}
\DeclareSIUnit{\dBW}{\deci\belwatt}

\usepackage{empheq}

\newtheorem{assump}{A.}

\usepackage{stackengine}
\def\delequal{\mathrel{\ensurestackMath{\stackon[1pt]{=}{\scriptstyle\Delta}}}}

\usepackage{algorithm,algcompatible}
\usepackage{algpseudocode}
\renewcommand{\COMMENT}[2][.7\linewidth]{%
  \leavevmode\hfill\makebox[#1][l]{//~#2}}
  
\usepackage[acronym]{glossaries}
\newacronym{cfo}{CFO}{carrier frequency offset}
\newacronym{emse}{EMSE}{excess mean-squared error}
\newacronym{fd}{FD}{full-duplex}
\newacronym{fir}{FIR}{finite impulse response}
\newacronym{fo-lms}{FO-LMS}{frequency offsets-compensated least mean squares}
\newacronym{kic}{KIC}{known-interference cancellation}
\newacronym{lms}{LMS}{least mean squares}
\newacronym{mse}{MSE}{mean-squared error}
\newacronym{nlms}{NLMS}{normalized least mean squares}
\newacronym{ocxo}{OCXO}{oven-controlled crystal oscillator}
\newacronym{ofdm}{OFDM}{orthogonal frequency division multiplexing}
\newacronym{rls}{RLS}{recursive least squares}
\newacronym{sdr}{SDR}{software-defined radio}
\newacronym{sfo}{SFO}{sampling frequency offset}
\newacronym{sgd}{SGD}{stochastic gradient descent}
\newacronym{snr}{SNR}{signal-to-noise ratio}
\newacronym{src}{SRC}{sampling rate conversion}
\newacronym{tcxo}{TCXO}{temperature-compensated crystal oscillator}
\newacronym{vss-fo-lms}{VSS-FO-LMS}{variable step sizes frequency offsets-compensated least mean squares}
\newacronym{vss-nlms}{VSS-NLMS}{variable step size normalized least mean squares}

\begin{document}

\title{A Variable Step Sizes\\Frequency Offsets-Compensated\\Least Mean Squares Algorithm}

\author{Karel~P{\"a}rlin,
        Aaron~Byman,
        Tommi~Meril{\"a}inen, and
        Taneli~Riihonen% <-this % stops a space
\thanks{This work has been submitted to the IEEE for possible publication. Copyright may be transferred without notice, after which this version may no longer be accessible.}% <-this % stops a space
\thanks{K.~P{\"a}rlin and T.~Riihonen are with Tampere University, Faculty of Information Technology and Communication Sciences, Korkeakoulunkatu~1, 33720 Tampere, Finland (e-mail: karel.parlin@tuni.fi; taneli.riihonen@tuni.fi).}% <-this % stops a space
\thanks{A.~Byman and T.~Meril{\"a}inen are with Bittium Wireless, Ritaharjuntie~1, 90590 Oulu, Finland (e-mail: aaron.byman@bittium.com; tommi.merilainen@bittium.com).}% <-this % stops a space
}

% The paper headers
\markboth{}%
{P{\"a}rlin \MakeLowercase{\textit{et al.}}: A Variable Step Sizes Frequency Offsets-Compensated Least Mean Squares Algorithm}

\maketitle

\begin{abstract}
\Gls{fo-lms} algorithm is a generic method for estimating a wireless channel under carrier and sampling frequency offsets when the transmitted signal is beforehand known to the receiver. The algorithm iteratively and explicitly adjusts its estimates of the channel and frequency offsets using stochastic gradient descent-based rules and the step sizes of these rules determine the learning rate and stability of the algorithm. Within the stability conditions, the choice of step sizes reflects a trade-off between the algorithm's ability to react to changes in the channel and the ability to minimize misadjustments caused by noise. This paper provides theoretical expressions to predict and optimize the tracking and misadjusment errors of \gls{fo-lms} when estimating channels and frequency offsets with known time-varying characteristics. This work also proposes a method to adjust the \gls{fo-lms}'s step sizes based on the algorithm's performance when the time-varying characteristics are not known, which is more often the case in practice. Accuracy of the expressions and performance of the proposed variable step sizes algorithm are studied through simulations.
\end{abstract}

\begin{IEEEkeywords}
Adaptive filtering, channel estimation, frequency offset, mean-squared error, steady-state analysis.
\end{IEEEkeywords}

\IEEEpeerreviewmaketitle

\section{Introduction}

\IEEEPARstart{T}{he} \gls{fo-lms} adaptive filter~\cite{parlin2023estimating} is a robust and low-complexity algorithm for estimating the propagation channel as well as the carrier and sampling frequency offsets between a radio transmitter and receiver. It has been demonstrated to be especially useful in \gls{kic}~\cite{parlin2023known, parlin2024distributed, parlin2024wideband, parlin2025high, parlin2025rayleigh} but, owing to its simplicity and generic structure, the algorithm is also of potential use in other signal processing applications, such as wireless communications, distributed beamforming, and bistatic radars~\cite{parlin2023estimating}. The algorithm works by iteratively and explicitly updating its estimates of the channel impulse response, carrier frequency offset, and sampling frequency offset between a transmitter and receiver based on \gls{sgd} rules according to user-defined step sizes and a cost function, which in the \gls{fo-lms}'s case is the instantaneous difference between the estimated received signal and the actual received signal. The step sizes control the algorithm's adaptation rate and in nonstationary conditions their selection becomes a trade-off between tracking ability and misadjustment. There will thus exist optimal step sizes that minimize the total estimation error, which is the sum of the tracking and adjustment errors.

Performance of classical adaptive filters in nonstationary conditions has been thoroughly studied. Some approaches have considered the impact of a cyclic nonstationarity only~\cite{rupp1998lms, xiao2005new}, while the energy conservation relation framework has been proposed to consider the impact of both cyclic and random nonstationarities in general adaptive filters~\cite{yousef2000generalized, yousef2001unified, yousef2002ability} and normalized adaptive filters~\cite{moinuddin2003tracking0, moinuddin2003tracking1}. The results in these studies highlighted the vulnerability of classical adaptive filters to cyclic nonstationarities and motivated the development of the \gls{fo-lms} adaptive filter in the first place. The performance analysis of \gls{fo-lms} has since demonstrated the benefit of explicitly estimating the cyclic nonstationarities as separate parameters~\cite{parlin2023estimating}.

However, the existing performance analyzes~\cite{parlin2023estimating, xiao2005new, yousef2000generalized, yousef2001unified, yousef2002ability, moinuddin2003tracking0, moinuddin2003tracking1} have assumed the cause of the cyclic nonstationarities to be static or time-invariant. In practice, the oscillators in radios exhibit both short-term and long-term instabilities that result in time-varying cyclic nonstationarities~\cite{bednarz2024frequency}. For example, the one second Allan deviation~\cite{riley2008handbook} of \glspl{tcxo} is typically around $10^{-9}$ and $10^{-10}$~\cite{bruggemann2006chip, lei2019time}. This instability results in notable carrier and sampling frequency offset drifts, and understanding the impact of these drifts on the \gls{fo-lms} algorithm is essential in finding its optimal step sizes.

Furthermore, in practice the time-varying characteristic of the cyclic and random nonstationarities are rarely precisely known, and it would be convenient to be able to use the \gls{fo-lms} algorithm without having to manually control its step sizes anyway. Ideally the algorithm would have the means to self-adapt the step sizes close to their optimal values based on some assessment. In that vein, numerous self-adapting step size techniques have been proposed to enhance classical adaptive filters~\cite{kwong1992variable, krstajic2002approach, shin2004variable, gerzaguet2018multiplicative, benesty2006nonparametric, huang2011new, paleologu2015overview} and step size adjustment in \gls{sgd}-based methods has been extensively covered in machine learning~\cite{duchi2011adaptive, zeiler2012adadelta, schaul2013no, kingma2014adam}. While not directly applicable herein, these works provide valuable insight on the possibilities of self-adapting the step sizes of \gls{fo-lms}.

\newpage
The purpose of this paper is twofold. Firstly, to present theoretical expressions for estimating the \gls{fo-lms} performance under time-varying cyclic and random nonstationarities, therefore allowing, e.g., instead of exhaustive numerical search, to calculate the optimal step sizes for the algorithm when all the system parameters are known. Secondly, to provide a method for the \gls{fo-lms} to self-adjust its step sizes, therefore removing the need to know all of the system parameters and consequently simplifying practical applications. We show that the theoretical estimates of the \gls{fo-lms} performance closely match with simulation results and that the theoretical optimal step sizes effectively minimize the sum of the tracking and adjustment errors. We also show that the proposed variable step size algorithm performs well in comparison to \gls{fo-lms} with optimal step sizes.

The rest of this article is organized as follows. In Section~\ref{sec:steady_state_analysis}, the system model is presented and expressions for estimating the steady-state performance as well as optimal step sizes of \gls{fo-lms} under time-varying channel and frequency offsets are derived. In Section~\ref{sec:variable_step_sizes}, a variable step sizes version of the \gls{fo-lms} algorithm is proposed. Section~\ref{sec:results} analyzes the accuracy of the theoretical performance and optimal step size expressions by comparison to simulations as well as evaluates the performance of the proposed \gls{vss-fo-lms}. Finally, conclusions of the study are given in Section~\ref{sec:conclusion}.

{\em Notation:} In this article, small boldface letters are used to denote vectors and capital boldface letters to denote matrices, e.g., $\mathbf{w} \in \mathbb{C}^{M \times 1} = [w_1 \enspace w_2 \enspace \cdots \enspace w_M]^T$ and $\mathbf{A} \in \mathbb{C}^{M \times N}$. The identity matrix is denoted by $\mathbf{I}$ and a zero vector is denoted by the boldface letter $\mathbf{0}$, both with dimensions compatible to each context. By default, vectors are column vectors. The iteration index is placed as a subscript for vectors and between parentheses for scalars, e.g., $\mathbf{w}_n$ and $v(n)$. Transpose, Hermitian transpose, and complex conjugate are denoted by $(\cdot)^T$, $(\cdot)^H$, and $(\cdot)^*$, respectively. Absolute-value norm is denoted by $\lvert\,\cdot\,\rvert$ and square norm by $\lVert\,\cdot\,\rVert$. Lastly, the real and imaginary operators are denoted by $\Re\{\cdot\}$ and $\Im\{\cdot\}$, respectively.

\section{Steady-State Analysis}
\label{sec:steady_state_analysis}

\subsection{System Model}
\label{ssec:theory_system_model}

\begin{figure}[t!]
\centerline{\includegraphics{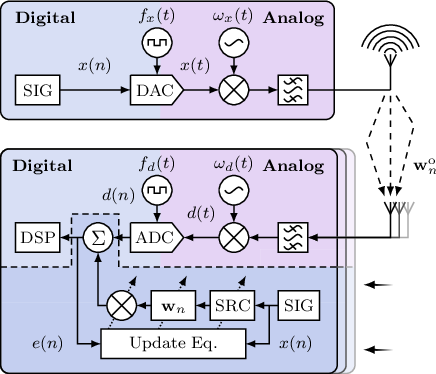}}
\caption{System model for estimating a wireless channel under carrier and sampling frequency offsets.}
\label{fig:system_model}
\end{figure}

The system model considered in this work is as follows and as illustrated in Fig.~\ref{fig:system_model}.  Let $x(n)$ denote a discrete-time complex-valued signal that is transmitted over a channel with a finite impulse response
\begin{equation}
    \mathbf{w}^\mathrm{o}_n = [w^\mathrm{o}_1(n) \quad w^\mathrm{o}_2(n) \quad \cdots \quad w^\mathrm{o}_{M}(n)]
\end{equation}
of order $M$ that varies in time according to a first order autoregressive process so that
\begin{equation}
    \mathbf{w}^\mathrm{o}_n = \mathbf{w}^\mathrm{o} + \bm{\theta}_n,
\end{equation}
where
\begin{equation}
    \bm{\theta}_{n+1} = \alpha \bm{\theta}_n + \mathbf{q}_n
\end{equation}
with $0 \leq \lvert \alpha \rvert < 1$ and $\mathbf{q}_n$ is a zero-mean stationary vector process with a positive-definite covariance matrix $\mathbf{Q} = E(\mathbf{q}_n\mathbf{q}^*_n)$ and $\mathrm{Tr}(\mathbf{Q}) = M \sigma^2_\mathrm{q}$.

Denoting the regressor of the channel by
\begin{equation}
    \mathbf{x}_n = [x(n) \quad x(n-1) \quad \cdots \quad x(n-M+1)],
\end{equation}
the received signal becomes
\begin{equation}\label{eq:d}
    d(n) = (\mathbf{w}^\mathrm{o}_n)^H \mathbf{y}^\mathrm{o}_n e^{j\phi^\mathrm{o}(n)} + v(n),
\end{equation}
where $v(n)= g(n) + s(n)$ is the measurement noise that consists of zero-mean Gaussian noise $g(n)$ with variance $\sigma^2_\mathrm{g}$ and potentially of a background signal $s(n)$, altogether having variance $\sigma^2_\mathrm{v}$. The term $e^{j\phi^\mathrm{o}(n)}$ accounts for the noise and drift in the carrier frequency offset based on the three-state clock model~\cite{zucca2005clock, kim2011tracking, david2015modeling} such that
\begin{align}
    \phi^\mathrm{o}(n+1) &= \phi^\mathrm{o}(n) + \epsilon^\mathrm{o}(n) T_\mathrm{s} + \omega_\upphi(n),\\
    \epsilon^\mathrm{o}(n+1) &= \epsilon^\mathrm{o}(n) + \omega_\upepsilon(n) + \kappa,
\end{align}
where $\phi^\mathrm{o}(n)$ and $\epsilon^\mathrm{o}(n)$ correspond to the carrier phase and frequency offset, respectively, and $T_\mathrm{s}$ is the nominal sampling period. The two zero-mean Gaussian random processes, $\omega_\upphi(n)$ and $\omega_\upepsilon(n)$, represent the random-walk phase noise and random-walk carrier frequency offset drift~\cite{giorgi2011performance} with variances $\sigma^2_\upphi$ and $\sigma^2_\upepsilon$, respectively, and the constant term $\kappa$ represents the linear carrier frequency offset drift. The received digitized baseband signal is a resampled version of the known transmitted signal so that
\begin{equation}
    \mathbf{y}^\mathrm{o}_n = [y^\mathrm{o}(n) \quad y^\mathrm{o}(n-1) \quad \cdots \quad y^\mathrm{o}(n-M+1)]
\end{equation}
with
\begin{equation}\label{eq:sample_rate_offset}
    y^\mathrm{o}(n) = x\left(n T_\mathrm{s} + \beta^\mathrm{o}(n)\right),
\end{equation}
where the term $\beta^\mathrm{o}(n)$ accounts for the jitter and drift in the sampling frequency offset based on the three-state clock model so that
\begin{align}
    \beta^\mathrm{o}(n+1) &= \beta^\mathrm{o}(n) + \sfrac{1}{\eta^\mathrm{o}(n)} + \omega_\upbeta(n),\\
    \eta^\mathrm{o}(n+1) &= \eta^\mathrm{o}(n) + \omega_\upeta(n) + \rho,
\end{align}
where $\beta^\mathrm{o}(n)$ and $\eta^\mathrm{o}(n)$ correspond to the sampling jitter and frequency offset, respectively. The two zero-mean Gaussian random processes, $\omega_\upbeta(n)$ and $\omega_\upeta(n)$, represent the random-walk sampling jitter and random-walk frequency offset drift with variances $\sfrac{T_\mathrm{s}\sigma^2_\upbeta}{2\pi}$ and $\sfrac{\sigma^2_\upeta}{2\pi}$, respectively, and the constant term $\rho$ represents the linear sampling frequency offset drift.

The \gls{fo-lms} algorithm provides estimates $\mathbf{w}_n$, $\epsilon(n)$, $\phi(n)$, $\eta(n)$, and $\beta(n)$ of the channel and clock offsets relying on the gradient of the instantaneous error
\begin{equation}\label{eq:e}
    e(n) = d(n) - \mathbf{w}^H_{n} \mathbf{y}_n e^{j \phi(n)}
\end{equation}
by iteratively updating its parameter estimates using the gradient descent equations
\begin{subequations}\label{eq:updates}
\begin{align}\label{eq:cir_update}
 \mathbf{w}_{n+1} =\ & \mathbf{w}_{n} + \mu_{\mathrm{w}} \mathbf{y}_n e^{j \phi(n) } e^{*}(n), \\\label{eq:cfo_update}
 \epsilon(n+1) =\ & \epsilon(n) + \mu_{{\upepsilon}} \Im{\left\{ \mathbf{w}^H_{n} \mathbf{y}_n e^{j \phi(n) } e^{*}(n) \right\}}, \\\label{eq:sfo_update}
 \eta(n+1) =\ & \eta(n) + \mu_{{\upeta}} \Re{\left\{ \mathbf{w}^H_{n} \mathbf{y}'_n  e^{j \phi(n) }e^{*}(n) \right\}},
\end{align}
\end{subequations}
where $\mathbf{y}'_n$ is the derivative of $\mathbf{y}_n$, and $\mu_{\mathrm{w}}$, $\mu_{{\upepsilon}}$, and $\mu_{{\upeta}}$ are the fixed channel, carrier, and sampling frequency offset estimation step sizes, respectively.

\subsection{Steady-State Performance}
\label{ssec:theory_steady_state_tracking_performance}

Due to the measurement noise and time-varying system model, the estimated parameters will inevitably differ from the true parameters. These differences can be denoted as
\begin{subequations}\label{eq:parameter_differences}
\begin{align}\label{eq:cir_difference}
    \tilde{\mathbf{w}}_{n} =\ & \hat{\mathbf{w}}^\mathrm{o}_n - \mathbf{w}_n,\\\label{eq:cfo_difference}
    \tilde{\epsilon}(n) = \epsilon^\mathrm{o}(n) - \epsilon(n)&\text{ and }\tilde{\phi}(n) = \phi^\mathrm{o}(n) - \phi(n),\\
    \tilde{\eta}(n) = \eta^\mathrm{o}(n) - \eta(n)&\text{ and }\tilde{\beta}(n) = \beta^\mathrm{o}(n) - \beta(n),
\end{align}
\end{subequations}
where
\begin{equation}\label{eq:w^o_n}
    \hat{\mathbf{w}}^\mathrm{o}_n \delequal \mathbf{T}_n \mathbf{w}^\mathrm{o}_n e^{j\tilde{\phi}(n)}
\end{equation}
and $\mathbf{T}_n \in \mathbb{C}^{M \times M}$ is an arbitrary time-shift matrix that, when multiplied with $\mathbf{x}_n$, delays the signal $\mathbf{x}_n$ by $\tilde{\beta}(n)$.

It is reasonable to assume that in steady state the parameter errors vary around zero so that $\tilde{\epsilon}(n) \ll 1$ and $\tilde{\eta}(n) \ll \sfrac{1}{f_\mathrm{max}}$ with $f_\mathrm{max}$ being the maximum frequency component of $x(n)$. The \textit{a priori} estimation errors relating to each of the update equations in (\ref{eq:updates}) can be then written as
\begin{subequations}\label{eq:a_priori_errors}
\begin{align}
    e^\mathrm{a}_\mathrm{w}(n) =\ & \tilde{\mathbf{w}}^{H}_{n} \mathbf{y}_n e^{j\phi(n)},\\
    e^\mathrm{a}_\mathrm{\upepsilon}(n) =\ & \mathbf{w}_{n}^{H} \mathbf{y}_n \tilde{\epsilon}(n) e^{j\phi(n)},\\
    e^\mathrm{a}_\mathrm{\upeta}(n) =\ & \mathbf{w}_{n}^{H} \mathbf{y}'_n \tilde{\eta}(n) e^{j\phi(n)},
\end{align}
\end{subequations}
so that the total instantaneous error in (\ref{eq:e}) is
\begin{equation}\label{eq:e_total_a_priori}
    e(n) = e^\mathrm{a}_\mathrm{w}(n) + e^\mathrm{a}_\mathrm{\upepsilon}(n) + e^\mathrm{a}_\mathrm{\upeta}(n) + v(n).
\end{equation}
Assuming furthermore that in steady state the noise sequence $v(n)$ and \textit{a priori} errors in (\ref{eq:a_priori_errors}) are stationary and statistically independent, the \gls{mse} is given by
\begin{equation}
    \mathrm{MSE} = \zeta + \sigma^2_v = \zeta_\mathrm{w} + \zeta_\upepsilon + \zeta_\upeta + \sigma^2_v,
\end{equation}
where $\zeta$ is the total \gls{emse} and
\begin{subequations}
\begin{align}
    \zeta_\mathrm{w} =\ & \lim_{n\to\infty} E\left[ |e^\mathrm{a}_\mathrm{w}(n)|^2 \right],\\
    \zeta_\upepsilon =\ & \lim_{n\to\infty} E\left[ |e^\mathrm{a}_\mathrm{\upepsilon}(n)|^2 \right],\\
    \zeta_\upeta =\ & \lim_{n\to\infty} E\left[ |e^\mathrm{a}_\mathrm{\upeta}(n)|^2 \right],
\end{align}
\end{subequations}
are the individual \glspl{emse} related to the update equations (\ref{eq:cir_update}), (\ref{eq:cfo_update}), and (\ref{eq:sfo_update}), respectively.

Employing the energy conservation relation~\cite{yousef2002ability}, it is shown in Appendix~\ref{ap:emse} that for sufficiently small step sizes and for Gaussian white input signals ($\mathbf{R} = \sigma^2_\mathrm{x}\mathbf{I}$) the individual steady-state \glspl{emse} can be approximated as
\begin{subequations}\label{eq:emse_all_gaussian}
% \begin{empheq}[box=\widefbox]{align}\label{eq:emse_cir_gaussian}
\begin{align}\label{eq:emse_cir_gaussian}
    \zeta_\mathrm{w} &\approx \frac{\mu_\mathrm{w} M \sigma^2_\mathrm{x} \sigma^2_\mathrm{v}}{2}  
    + \frac{M\sigma^2_\mathrm{q}}{2\mu_\mathrm{w}}\\
    &\qquad\ + \frac{\mu_\upepsilon \lVert\mathbf{w}^\mathrm{o}\rVert^2 \sigma^2_\mathrm{v}}{4\mu_\mathrm{w}}
    + \frac{\mu_\upeta \lVert\mathbf{w}^\mathrm{o}\rVert^2 \sigma^2_\mathrm{v}}{2\mu_\mathrm{w}}\nonumber\\
    &\qquad\ + \frac{ \lVert\mathbf{w}^\mathrm{o}\rVert^2 \sigma^2_\upphi }{2 \mu_\mathrm{w}}
    + \frac{
    \lVert\mathbf{w}^\mathrm{o}\rVert^2 \sigma^2_\upbeta}{2 \mu_\mathrm{w} T_\mathrm{s} },
    \nonumber\\\label{eq:emse_cfo_gaussian}
    \zeta_\upepsilon &\approx
    \frac{\mu_\upepsilon \sigma^2_\mathrm{x} \lVert\mathbf{w}^\mathrm{o}\rVert^2 \sigma^2_\mathrm{v}}{2} +
    \frac{\sigma^2_\upepsilon T_\mathrm{s}^2}{2 \mu_\mathrm{w} \mu_\upepsilon \sigma^2_\mathrm{x}}\\
    &\qquad\ + \frac{\kappa^2 T_\mathrm{s}^2}{\mu^2_\upepsilon \sigma^2_\mathrm{x} \lVert\mathbf{w}^\mathrm{o}\rVert^2},\nonumber
    \\\label{eq:emse_sro_gaussian}
    \zeta_\upeta &\approx
    \mu_\upeta \sigma^2_\mathrm{x} \lVert\mathbf{w}^\mathrm{o}\rVert^2 \sigma^2_\mathrm{v}
     + \frac{
     \sigma^2_\upeta T_\mathrm{s}^2}{2 \mu_\mathrm{w} \mu_\upeta \sigma^2_\mathrm{x}}
     \nonumber\\
    &\ + \frac{ \rho^2 T_\mathrm{s}^2 }{2 \left(2 + \sfrac{2}{M} \right) \mu^2_\upeta \sigma^2_\mathrm{x} \lVert\mathbf{w}^\mathrm{o}\rVert^2}
     + \frac{ \mu_\mathrm{w} \lVert\mathbf{w}^\mathrm{o}\rVert^2 \sigma^2_\upbeta}{ 2 \mu_\upeta T_\mathrm{s} }.
\end{align}
% \end{empheq}
\end{subequations}
At large step sizes, and as long as the algorithm remains in its steady state, accuracy of the \gls{emse} estimation is improved by using the complete expressions in \eqref{eq:emse_all_gaussian_complete}. For example, the estimated \glspl{emse} in this paper (i.e., in Section~\ref{sec:results}) are calculated using \eqref{eq:emse_all_gaussian_complete}. However, equations in \eqref{eq:emse_all_gaussian} are used in the following optimal step size derivations leading to more compact expressions.

\subsection{Optimal Step Sizes}
\label{ssec:theory_optimal_step_sizes}

From \eqref{eq:emse_all_gaussian} it is evident that the individual \glspl{emse} of the update equations are not monotonically increasing functions of the corresponding step sizes. Instead, the \glspl{emse} are composed of multiple terms with differing impacts from each of the step sizes. The first terms on the right-hand sides in \eqref{eq:emse_all_gaussian} denote the misadjustments that decrease as the corresponding step sizes are decreased (i.e., as the algorithm becomes less susceptible to noise). The rest of the terms denote the tracking errors that decrease as the corresponding step sizes are increased (i.e., as the algorithm can more rapidly react to changes in the true parameters).

Specifically, the second, fifth and sixth term in \eqref{eq:emse_cir_gaussian} denote the tracking errors with regards to the random channel perturbations, random-walk phase noise, and random-walk sampling jitter, respectively. The third and fourth term in \eqref{eq:emse_cir_gaussian} denote the tracking errors of the self-induced nonstationarities that the \gls{fo-lms} inherently produces and these decrease as the step size $\mu_\mathrm{w}$ is increased and as the step sizes $\mu_\upepsilon$ and $\mu_\upeta$ are decreased (i.e., as the variance of the self-induced nonstationarities is reduced). Finally, the second and third terms in \eqref{eq:emse_cfo_gaussian} and \eqref{eq:emse_sro_gaussian} denote the tracking errors with regards to the random-walk and linear drift of the frequency offsets. The fourth term in \eqref{eq:emse_sro_gaussian} denotes the misadjustment due to the increase in the noise floor caused by the accumulated sampling jitter and its inadequate tracking at small $\mu_\upeta$.

It is evident that there exists a set of values $\mu^\mathrm{opt}_{\mathrm{w}}$, $\mu^\mathrm{opt}_{{\upepsilon}}$, and $\mu^\mathrm{opt}_{{\upeta}}$ that minimizes the \gls{emse}. These optimal values can be obtained by taking the partial derivatives of the total \gls{emse} with regards to the step sizes and then solving the partial derivatives for the step sizes resulting in
\begin{subequations}\label{eq:optimal_step_sizes}
% \begin{empheq}[box=\widefbox]{align}
\begin{align}
    \mu^\mathrm{opt}_{\mathrm{w}} &= \biggl\{ \frac{1}{
    M \sigma^2_\mathrm{v} \sigma^2_\mathrm{x}
    + \frac{ \lVert\mathbf{w}^\mathrm{o}\rVert^2 \sigma^2_\upphi}{\mu_\upeta T_\mathrm{s}}
    }\\
    &\quad\ \times \biggl(
    M \sigma^2_\mathrm{q}
    + \frac{\mu^\mathrm{opt}_\upepsilon \lVert\mathbf{w}^\mathrm{o}\rVert^2 \sigma^2_\mathrm{v}}{2} + \mu^\mathrm{opt}_\upeta \lVert\mathbf{w}^\mathrm{o}\rVert^2 \sigma^2_\mathrm{v}\nonumber\\
    &\ + 
    \frac{\lVert\mathbf{w}^\mathrm{o}\rVert^2 \sigma^2_\upbeta}{T_\mathrm{s}}
    + \lVert\mathbf{w}^\mathrm{o}\rVert^2 \sigma^2_\upphi
    + \frac{\sigma^2_\upepsilon T^2_\mathrm{s}}{ \mu_\upepsilon \sigma^2_\mathrm{x} }
    + \frac{
    \sigma^2_\upeta T^2_\mathrm{s}}{ \mu_\upeta \sigma^2_\mathrm{x} } \biggr) \biggr\}^{\!1/2},\nonumber\\\label{eq:mu_cfo_optimal}
    \mu^\mathrm{opt}_{{\upepsilon}} &\approx \sqrt{\frac{2 \sigma^2_\upepsilon T^2_\mathrm{s} }{ \lVert\mathbf{w}^\mathrm{o}\rVert^2 \sigma^2_\mathrm{v} \sigma^2_\mathrm{x} \left(2 \mu_\mathrm{w} \sigma^2_\mathrm{x} + 1 \right)}}\\
    &\qquad\ + \sqrt[3]{\frac{ 8 \mu^\mathrm{opt}_\mathrm{w} \kappa^2 T^2_\mathrm{s}}{2 \lVert\mathbf{w}^\mathrm{o}\rVert^4 \mu^\mathrm{opt}_\mathrm{w} \sigma^2_\mathrm{v} \sigma^4_\mathrm{x} + \lVert\mathbf{w}^\mathrm{o}\rVert^4 \sigma^2_\mathrm{v} \sigma^2_\mathrm{x}} },\nonumber\\\label{eq:mu_sfo_optimal}
    \mu^\mathrm{opt}_{{\upeta}} &\approx
    \sqrt{\frac{
    \sfrac{
    \lVert\mathbf{w}^\mathrm{o}\rVert^2 \sigma^2_\upbeta \mu^2_\mathrm{w} \sigma^2_\mathrm{x}}{T_\mathrm{s}}
    % \pi
    + \sigma^2_\upeta T^2_\mathrm{s}}{ \lVert\mathbf{w}^\mathrm{o}\rVert^2 \sigma^2_\mathrm{v} \sigma^2_\mathrm{x} \left(2 \mu_\mathrm{w} \sigma^2_\mathrm{x} + 1 \right)}}\\
    &\qquad\ + \sqrt[3]{ \frac{ \mu^\mathrm{opt}_\mathrm{w} \rho^2 T^2_\mathrm{s}}{2 \lVert\mathbf{w}^\mathrm{o}\rVert^4 \mu^\mathrm{opt}_\mathrm{w} \sigma^2_\mathrm{v} \sigma^4_\mathrm{x} + \lVert\mathbf{w}^\mathrm{o}\rVert^4 \sigma^2_\mathrm{v} \sigma^2_\mathrm{x} } }.\nonumber
\end{align}
% \end{empheq}
\end{subequations}
Given the time-varying characteristics of the system parameters, the system of equations in \eqref{eq:optimal_step_sizes} can be solved using numerical optimizers, such as \cite{la2006spectral}, to provide step sizes that minimize the total \gls{emse} of the \gls{fo-lms} algorithm when it is in steady state. Despite the numerical optimization, the above equations provide several orders of magnitude reduction in computational effort in comparison to exhaustively testing for the optimality of parameters like required in the prior art.

To assist numerical algorithms in optimizing the system of equations in~\eqref{eq:optimal_step_sizes}, it can be beneficial to provide initial guesses for the optimal step sizes without considering the coupling of the individual \glspl{emse}. That is, by deriving an approximal optimal step size $\hat{\mu}^\mathrm{opt}_\mathrm{w}$ based on~\eqref{eq:emse_cir_gaussian} without the cross-terms so that
\begin{equation}\label{eq:mu_cir_optimal_approximal}
    \hat{\mu}^\mathrm{opt}_\mathrm{w} = 
    \sqrt{
    \frac{
    M \sigma^2_\mathrm{q}
    + \sfrac{ \lVert\mathbf{w}^\mathrm{o}\rVert^2 \sigma^2_\upbeta }{ T_\mathrm{s} }
    + \lVert\mathbf{w}^\mathrm{o}\rVert^2 \sigma^2_\upphi
    }{M \sigma^2_\mathrm{v} \sigma^2_\mathrm{x}}}.
\end{equation}
Then, by first calculating $\hat{\mu}^\mathrm{opt}_\mathrm{w}$ based on \eqref{eq:mu_cir_optimal_approximal}, $\mu^\mathrm{opt}_\upepsilon$ and $\mu^\mathrm{opt}_\upeta$ can be directly evaluated with \eqref{eq:mu_cfo_optimal} and \eqref{eq:mu_sfo_optimal}, respectively, using $\hat{\mu}^\mathrm{opt}_\mathrm{w}$ in place of $\mu^\mathrm{opt}_\mathrm{w}$. Relying on the resulting optimal step size approximations as starting points in numerical optimizers for solving \eqref{eq:optimal_step_sizes} will generally lead to a reduced number of required optimization iterations.

\section{Variable Step Sizes}
\label{sec:variable_step_sizes}

In this section, we propose a method to determine reasonable step sizes of the \gls{fo-lms} algorithm without explicit knowledge of the time varying characteristics of the system parameters.

\subsection{Derivation of Variable Step Sizes}

As highlighted by the previous section, the individual \glspl{emse} of the \gls{fo-lms} update equations are coupled and the selection of any one of the three step sizes not only affects the \gls{emse} of the update equation controlled by that step size but also the \glspl{emse} of the other update equations. Since the three update equations of \gls{fo-lms} operate over the same cost function, we propose to use that cost function as feedback to directly adjust the step size $\mu_\mathrm{w}$ that is directly linked to the update equation \eqref{eq:cir_update}, which in practical situations likely contributes the majority of the \gls{emse}. We then propose to adjust the remaining step sizes $\mu_\upepsilon$ and $\mu_\upeta$ depending on the already selected step size $\mu_\mathrm{w}$ and the time-averaged gradients of the remaining two update equations.

The equations for selecting the step sizes can be derived by first noting that the step sizes of the \gls{fo-lms} algorithm will be optimal if
\begin{equation}
    E\left[ \lvert e^\mathrm{p}(n) \rvert^2 \right] = \sigma^2_\mathrm{v} \qquad \forall n,
\end{equation}
where
\begin{equation}\label{eq:e_posteriori}
    e^\mathrm{p}(n) = d(n) - \mathbf{w}^H_{n+1} \mathbf{y}_n e^{j \phi(n)}
\end{equation}
is the \textit{a posteriori} error signal calculated at sample index $n$ using the updated parameter estimates $\mathbf{w}_{n+1}$, $\epsilon(n+1)$, and $\eta(n+1)$. Relying on the assumptions that the signal $x(n)$ is white Gaussian and that the channel update equation \eqref{eq:cir_update} contributes the majority of the total \gls{emse}, the expected value of the power of the \textit{a posteriori} error signal can be approximated as
\begin{equation}\label{eq:e_posteriori_expansion}
    E\left[ \lvert e^\mathrm{p}(n) \rvert^2 \right] \approx \left[ 1 - \mu_\mathrm{w}(n) M \sigma^2_\mathrm{y} \right]^2 \sigma^2_\mathrm{e}(n),
\end{equation}
where $\sigma^2_\mathrm{e}(n) = E\left[ \lvert e(n) \rvert^2 \right]$ is the power of the error signal. Then, taking the right-hand sides in \eqref{eq:e_posteriori} and \eqref{eq:e_posteriori_expansion} to be equal and solving for $\mu_\mathrm{w}(n)$ results in the nonparametric \gls{vss-nlms} algorithm~\cite{benesty2006nonparametric} with
\begin{equation}\label{eq:mu_cir_adjusted}
    \mu_\mathrm{w}(n) = \frac{1}{\mathbf{y}^H_n\mathbf{y}_n} \left[1 - \frac{\sigma_\mathrm{v}}{\sigma_\mathrm{e}(n)}\right].
\end{equation}

When the carrier and sampling frequency offsets do not change over time and the \gls{fo-lms} algorithm has reached a steady state with accurate estimates of the carrier and sampling frequency offsets, the gradients
\begin{subequations}
\begin{align}\label{eq:gradient_cfo}
    \Delta_\upepsilon(n) &= \Im{\left\{ \mathbf{w}^H_{n} \mathbf{y}_n e^{j \phi(n) } e^{*}(n) \right\}}, \\\label{eq:gradient_sfo}
    \Delta_\upeta(n) &= \Re{\left\{ \mathbf{w}^H_{n} \mathbf{y}'_n  e^{j \phi(n) }e^{*}(n) \right\}},
\end{align}
\end{subequations}
in \eqref{eq:cfo_update} and \eqref{eq:sfo_update} will be zero-mean wide sense stationary processes. However, when the frequency offsets drift over time, these gradients will have non-zero means $\bar{\Delta}_\upepsilon(n)$ and $\bar{\Delta}_\upeta(n)$ that are proportional to the carrier and sampling frequency offset drifts as well as scaled by the corresponding step sizes $\mu_\upepsilon$ and $\mu_\upeta$.

Optimal values for $\mu_\upepsilon(n)$ and $\mu_\upeta(n)$ depend on both the random-walk and linear drift of the carrier and sampling frequency offsets. However, the random-walk and linear drift components are in practice difficult to separate. We propose to calculate the remaining two variable step sizes by assuming that the gradient follows the linear drift and demonstrate in the next section that this leads to a satisfactory result. Given that the channel update step size $\mu_\mathrm{w}(n)$ has been calculated using \eqref{eq:mu_cir_adjusted}, optimal values for the remaining step sizes $\mu_\upepsilon(n)$ and $\mu_\upeta(n)$ can then be calculated based on \eqref{eq:mu_cfo_optimal} and \eqref{eq:mu_sfo_optimal} as%
\begin{subequations}\label{eq:variable_fo_step_sizes}
\begin{align}
    \mu_\upepsilon(n) &= \sqrt[3]{\frac{ 8 \mu_\mathrm{w}(n) \left( \bar{\Delta}_\upepsilon(n) \bar{\mu}_\upepsilon(n) \right)^2}{2 \lVert\mathbf{w}^\mathrm{o}\rVert^4 \mu_\mathrm{w}(n) \sigma^2_\mathrm{v} \sigma^4_\mathrm{x} + \lVert\mathbf{w}^\mathrm{o}\rVert^4 \sigma^2_\mathrm{v} \sigma^2_\mathrm{x}} } , \\
    \mu_\upeta(n) &= \sqrt[3]{ \frac{ \mu_\mathrm{w}(n) \left( \bar{\Delta}_\upeta(n) \bar{\mu}_\upeta(n) \right)^2}{2 \lVert\mathbf{w}^\mathrm{o}\rVert^4 \mu_\mathrm{w}(n) \sigma^2_\mathrm{v} \sigma^4_\mathrm{x} + \lVert\mathbf{w}^\mathrm{o}\rVert^4 \sigma^2_\mathrm{v} \sigma^2_\mathrm{x} } },
\end{align}
\end{subequations}
where $\bar{\mu}_\upepsilon(n)$ and $\bar{\mu}_\upeta(n)$ are running averages of the step sizes $\mu_\upepsilon(n)$ and $\mu_\upeta(n)$.

\subsection{Practical Considerations}

In practice, the quantities $\sigma^2_\mathrm{e}(n)$, $\sigma^2_\mathrm{y}(n)$, $\bar{\Delta}_\upepsilon(n)$, and $\bar{\Delta}_\upeta(n)$ need to be measured. Furthermore, they need to be averaged over time, since the instantaneous error and gradient values are misleading due to the measurement noise and algorithm's misadjustments. The prevailing method for keeping track of the changes in measurement errors and system parameters in adaptive filters and neural network training methods is to use the exponentially weighted moving average~\cite{hinton2012neural}, a simple and robust approach. The aforementioned quantities can be straightforwardly tracked as exponentially weighted running averages such that
\begin{subequations}
\begin{align}\label{eq:moving_average_error}
    \sigma^2_\mathrm{e}(n) =\ & \lambda_\mathrm{e} \sigma^2_\mathrm{e}(n-1) + (1-\lambda_\mathrm{e})\lvert e(n)\rvert^2,\\\label{eq:moving_average_y}
    \sigma^2_\mathrm{y}(n) =\ & \lambda_\mathrm{y} \sigma^2_\mathrm{y}(n-1) + (1-\lambda_\mathrm{y})\lvert y(n)\rvert^2,\\\label{eq:moving_average_epsilon}
    \bar{\Delta}_\upepsilon(n) =\ & \lambda_\upepsilon \bar{\Delta}_\upepsilon(n-1)
    + (1-\lambda_\upepsilon) \Delta_\upepsilon(n),\\\label{eq:moving_average_eta}
    \bar{\Delta}_\upeta(n) =\ & \lambda_\upeta \bar{\Delta}_\upeta(n-1)
    + (1-\lambda_\upeta) \Delta_\upeta(n),
\end{align}
\end{subequations}
where $\lambda_\mathrm{e}$, $\lambda_\mathrm{y}$, $\lambda_\upepsilon$, and $\lambda_\upeta$ are the weighting coefficients that control the exponential decay. The step size averages $\bar{\mu}_\upepsilon(n)$ and $\bar{\mu}_\upeta(n)$ can be tracked as exponentially weighted moving averages too. However, a practical solution that avoids introducing additional forgetting factors is to use the average of the last $M$ step sizes so that $\bar{\mu}_\upepsilon(n) = \sfrac{1}{M} \sum^{n-1}_{i=n-M}\mu_\upepsilon(i)$ and $\bar{\mu}_\upeta(n) = \sfrac{1}{M} \sum^{n-1}_{i=n-M}\mu_\upeta(i)$.

Another aspect to consider in practice is the permanence of the measurement noise. In scenarios where the measurement noise is stable, it is sufficient to measure its variance $\sigma^2_\mathrm{v}$ during a calibration phase when $x(n)$ is not being transmitted. However, if the measurement noise variance changes noticeably after calibration, the calibrated measure of $\sigma^2_\mathrm{v}$ will lead to a poor choice of step sizes. This will almost certainly be the case in practice when the measurement noise includes a background signal(-of-interest) $s(n)$. To tackle this, the measurement noise variance can in practice be estimated with some accuracy and the choice of the optimal measurement noise estimator depends on the statistical properties of the signals involved~\cite{iqbal2008novel, paleologu2008variable, jung2017stabilization, strutz2019estimation, tiglea2022variable}. For the general case of white Gaussian signals, estimating the measurement noise based on the cross-correlation of the known and error signals can be a straightforward and reliable approach~\cite{strutz2019estimation}.

The reasoning behind that method relies on the cross-correlation vector between the known signal and the estimation error, herein being given by
\begin{align}
    \mathbf{R}_{\mathbf{y}\mathrm{e}}(n) =\ & E\left[\mathbf{y}_n e^{j \phi(n) } e^*(n)\right] \\\nonumber
     \delequal\ & \mathbf{R}_\mathbf{yy}\tilde{\mathbf{w}}_{n},
\end{align}
and the variance of the estimation error being
\begin{equation}
    \sigma^2_\mathrm{e}(n) = \tilde{\mathbf{w}}^H_{n} \mathbf{R}_\mathbf{yy} \tilde{\mathbf{w}}_{n} + \sigma^2_\mathrm{v}(n).
\end{equation}
Then, the estimate of the measurement noise is defined as
\begin{equation}\label{eq:noise_variance_estimate}
    \hat{\sigma}^2_\mathrm{v}(n) = \sigma^2_\mathrm{e}(n) - \frac{1}{\sigma^2_\mathrm{y}(n)} \hat{\mathbf{R}}_{\mathbf{y}\mathrm{e}}^{H}(n) \hat{\mathbf{R}}_{\mathbf{y}\mathrm{e}}(n),
\end{equation}
where an estimate of the cross-correlation between the resampled known signal $y(n)$ and the error signal $e(n)$ is
\begin{equation}\label{eq:cross_correlation_vector_estimation}
    \hat{\mathbf{R}}_{\mathbf{y}\mathrm{e}}(n) = \lambda_\mathrm{R} \hat{\mathbf{R}}_{\mathbf{y}\mathrm{e}}(n-1) + (1-\lambda_\mathrm{R})\mathbf{y}_n e^{j \phi(n) } e^*(n).
\end{equation}

The exponentially weighted running average of estimation error $\sigma^2_\mathrm{e}(n)$ may result in a lower magnitude than $\sigma^2_\mathrm{v}$ which would result in negative step sizes. When this happens, the step sizes should be simply set to zero. Furthermore, it can be beneficial to include regularization to avoid divisions by very small values by adding a small positive constant in the denominator of \eqref{eq:mu_cir_adjusted}. Also, if the time-varying characteristic of any of the system parameters are more or less known in advance, the step sizes can be limited to be within the expected sensible step size ranges, therefore avoiding any extreme step size values and improving the stability of the \gls{vss-fo-lms} algorithm. Similarly, if the measurement noise estimation in~\eqref{eq:noise_variance_estimate} is used because the environment is assumed to contain an unknown background signal $s(n)$ but the receiver noise floor $\sigma^2_\mathrm{g}$ is stable and known, then the noise variance estimate can be lower-bounded by the known $\sigma^2_\mathrm{g}$ in order to, again, improve the stability of the \gls{vss-fo-lms} algorithm.

The sampling frequency offset gradient in~\eqref{eq:gradient_sfo} includes a time derivative of the resampled signal vector. If the third derivative of $\mathbf{y}_n$ exists, then it is beneficial to use the centered first-order divided difference, which has an approximation error of order two~\cite[p.~172]{atkinson2005theoretical}, so that
\begin{equation}\label{eq:centered_difference}
    \mathbf{w}^H_{n} \mathbf{y}'_n \approx \frac{ \mathbf{w}^H_{n} \mathbf{y}_{n+1} - \mathbf{w}^H_{n} \mathbf{y}_{n-1}}{2(1+{\eta}(n))}.
\end{equation}
Alternatively, the first-order backward divided difference
\begin{equation}\label{eq:backward_difference}
    \mathbf{w}^H_{n} \mathbf{y}'_n \approx \frac{\mathbf{w}^H_{n}{\mathbf{y}}_{n} - \mathbf{w}^H_{n}{\mathbf{y}}_{n-1}}{1+{\eta}(n)}
\end{equation}
can be used, which does not require computation of $\mathbf{y}_{n+1}$ nor the existence of the third derivative, but has an approximation error of order one. In this work, the computationally more complex centered first-order divided difference~\eqref{eq:centered_difference} is used due to its higher accuracy providing a better foundation for comparing the estimated and simulated \glspl{emse}.

It should also be acknowledged that, like most \gls{sgd} algorithms, the \gls{fo-lms} may converge to a local minimum when dealing with multimodal error surfaces. In order to guarantee convergence to a global minimum, combining the \gls{fo-lms} algorithm with global search algorithms~\cite{ng1996genetic, gotmare2017swarm} or advanced \gls{sgd} techniques can be considered~\cite{loshchilov2017sgdr}. However, considering these in detail is out of the scope of this paper. Steps of the resulting \gls{vss-fo-lms} algorithm, considering the above-listed practical considerations, are listed as Algorithm~\ref{alg:vss-fo-lms}. Also, an open-source implementation is available as part of an adaptive filters toolkit.\footnote{\url{https://github.com/karel/gr-adapt}}

\begin{algorithm}
\caption{}\label{alg:vss-fo-lms}
\begin{algorithmic}[1]
\Procedure{VSS-FO-LMS}{$x$, $d$, $\mathbf{w}$, $\epsilon$, $\eta$, $M$, $\lambda_\mathrm{e}$,
\newline\phantom{-----------------------------------------} $\lambda_\upepsilon$, $\lambda_\upeta$, $\lambda_\mathrm{y}$, $\lambda_\mathrm{R}$, $\sigma^2_\mathrm{v}$}

\hskip\algorithmicindent
% \If{provided estimates $\mathbf{w}$, $\epsilon$, and $\eta$}
\State $\mathbf{w}_{0} \gets \mathbf{w}$ \COMMENT{\small zeros can be provided for}
\State $\epsilon(0) \gets \epsilon$ \COMMENT{\small the parameter estimates if no}
\State $\eta(0) \gets \eta$ \COMMENT{\small prior estimates are available}
% \Else
%     \State $\mathbf{w}_{0} \gets \mathbf{0}$, $\epsilon(0) \gets 0$, $\eta(0) \gets 0$
% \EndIf

\hskip\algorithmicindent
\State $\phi(1) \gets 0$, $t(1) \gets 0$
\State $\mu^\mathrm{min}_\mathrm{w} \gets 0$, $\mu^\mathrm{min}_\upepsilon \gets 0$, $\mu^\mathrm{min}_\upeta \gets 0$
\State $\mu^\mathrm{max}_\mathrm{w} \gets 1$, $\mu^\mathrm{max}_\upepsilon \gets 1$, $\mu^\mathrm{max}_\upeta \gets 1$
\State $\sigma^2_\mathrm{e}(0) \gets 1$, $\sigma^2_\mathrm{y}(0) \gets 0$, $\hat{\sigma}^2_\mathrm{v}(0) \gets 0$

\hskip\algorithmicindent

\For{$n \gets 1 \text{ to } N$}
\State $y(n) \gets x(t(n))$
\State $\mathbf{y}_n = [y(n) \quad y(n-1) \quad \cdots \quad y(n-M+1)]$

\hskip\algorithmicindent
\State $e(n) \gets d(n) - \mathbf{w}^H_{n} \mathbf{y}_n e^{j \phi(n)}$

\hskip\algorithmicindent
\State $\Delta_\upepsilon \gets \Im{\left\{ \mathbf{w}^H_{n} \mathbf{y}_n e^{j \phi(n) } e^{*}(n) \right\}}$
\State $\Delta_\upeta \gets \Re{\left\{ \mathbf{w}^H_{n} \mathbf{y}'_n  e^{j \phi(n) }e^{*}(n) \right\}}$

\hskip\algorithmicindent
\State $\sigma^2_\mathrm{e}(n) \gets \lambda_\mathrm{e} \sigma^2_\mathrm{e}(n-1) + (1-\lambda_\mathrm{e})\lvert e(n)\rvert^2$

\hskip\algorithmicindent
\If{$\sigma^2_\mathrm{v}$ is known}
    \State $\hat{\sigma}^2_\mathrm{v}(n) \gets \sigma^2_\mathrm{v}$
\Else
    \State $\sigma^2_\mathrm{y}(n) \gets \lambda_\mathrm{y} \sigma^2_\mathrm{y}(n-1) + (1-\lambda_\mathrm{y})\lvert y(n)\rvert^2$
    \State $\hat{\mathbf{R}}_{\mathbf{y}\mathrm{e}}(n) \gets \lambda_\mathrm{R} \hat{\mathbf{R}}_{\mathbf{y}\mathrm{e}}(n-1)$
    
    \hskip\algorithmicindent\hskip\algorithmicindent\hskip\algorithmicindent\hskip\algorithmicindent\hskip\algorithmicindent\hskip\algorithmicindent $ + (1-\lambda_\mathrm{R})\mathbf{y}_n e^{j \phi(n) } e^*(n)$
    \State $\hat{\sigma}^2_\mathrm{v}(n) \gets \sigma^2_\mathrm{e}(n) - \sfrac{1}{\sigma^2_\mathrm{y}(n)} \hat{\mathbf{R}}_{\mathbf{y}\mathrm{e}}^{H}(n) \hat{\mathbf{R}}_{\mathbf{y}\mathrm{e}}(n)$
\EndIf

\hskip\algorithmicindent
\State $\bar{\Delta}_\upepsilon(n) \gets \lambda_\upepsilon \bar{\Delta}_\upepsilon(n-1)+ (1-\lambda_\upepsilon) \Delta_\upepsilon$
\State $\bar{\Delta}_\upeta(n) \gets \lambda_\upeta \bar{\Delta}_\upeta(n-1) + (1-\lambda_\upeta) \Delta_\upeta$
\State $\bar{\mu}_\upepsilon(n) = \sfrac{1}{M} \sum^{n-1}_{i=n-M}\mu_\upepsilon(i)$
\State $\bar{\mu}_\upeta(n) = \sfrac{1}{M} \sum^{n-1}_{i=n-M}\mu_\upeta(i)$

\hskip\algorithmicindent
\State $\mu_\mathrm{w}(n) \gets \frac{1}{\mathbf{y}^H_n\mathbf{y}_n} \left[1 - \frac{\hat{\sigma}_\mathrm{v}(n)}{\sigma_\mathrm{e}(n)}\right]$
\\[]
\State $\mu_\upepsilon(n) \gets \sqrt[3]{\frac{ 8 \mu_\mathrm{w}(n) \left( \bar{\Delta}_\upepsilon(n) \bar{\mu}_\upepsilon(n) \right)^2}{\lVert\mathbf{w}_{n}\rVert^4 \hat{\sigma}^2_\mathrm{v}(n) \sigma^2_\mathrm{y}(n) \left[2 \mu_\mathrm{w}(n) \sigma^2_\mathrm{y}(n) + 1 \right] } }$
\\[]
\State $\mu_\upeta(n) \gets \sqrt[3]{ \frac{ \mu_\mathrm{w}(n) \left( \bar{\Delta}_\upeta(n) \bar{\mu}_\upeta(n) \right)^2}{\lVert\mathbf{w}_{n}\rVert^4 \hat{\sigma}^2_\mathrm{v}(n) \sigma^2_\mathrm{y}(n) \left[2 \mu_\mathrm{w}(n) \sigma^2_\mathrm{y}(n) + 1 \right] } }$

\hskip\algorithmicindent
\State $\mu_\mathrm{w}(n) \gets \max{ \left( \min{\left(\mu_\mathrm{w}(n), \mu^\mathrm{max}_\mathrm{w}\right)},  \mu^\mathrm{min}_\mathrm{w}\right) }$
\State $\mu_\upepsilon(n) \gets \max{ \left( \min{\left(\mu_\upepsilon(n), \mu^\mathrm{max}_\upepsilon\right)},  \mu^\mathrm{min}_\upepsilon\right) }$
\State $\mu_\upeta(n) \gets \max{ \left( \min{\left(\mu_\upeta(n), \mu^\mathrm{max}_\upeta\right)},  \mu^\mathrm{min}_\upeta\right) }$

\hskip\algorithmicindent
\State $\mathbf{w}_{n+1} \gets \mathbf{w}_{n} + \mu_\mathrm{w}(n) \mathbf{y}_n e^{j \phi(n) } e^{*}(n)$
\State $\epsilon(n+1) \gets \epsilon(n) + \mu_\upepsilon(n) \Delta_\upepsilon$
\State $\eta(n+1) \gets \eta(n) + \mu_\upeta(n) \Delta_\upeta$

\hskip\algorithmicindent
\State $\phi(n+1) \gets \phi(n) + \epsilon(n+1)$
\State $t(n+1) \gets t(n) + (1 + \eta(n+1))$

\EndFor
\EndProcedure
\end{algorithmic}
\end{algorithm}

\section{Results}
\label{sec:results}

In order to verify the theoretical expressions derived for \gls{fo-lms} in Section~\ref{sec:steady_state_analysis} and to evaluate the \gls{vss-fo-lms} algorithm proposed in Section~\ref{sec:variable_step_sizes}, performance of the two algorithms is herein studied through simulations. The simulations were carried out using white Gaussian noise as the known signal $x(n)$ with variance $\sigma^2_\mathrm{x}=0$~\SI{}{\dBW}. The measurement noise $v(n)$ was also simulated as white Gaussian but with variance $\sigma^2_\mathrm{v}=-60$ \SI{}{\dBW}. The carrier frequency offset was taken to be \SI{100}{\hertz}, the transmitter sampling frequency to be \SI{1}{\mega\hertz}, and the receiver sampling frequency to be that of the transmitter with a \SI{1}{\hertz} offset.

The known signal was oversampled with a factor of two so as to limit the effect of interpolation errors on the simulated \gls{emse} and to facilitate calculation of the derivative of the known signal vector in~\eqref{eq:sfo_update} using the centered first-order divided difference. The channel was simulated with an impulse response length of $M = 5$, autoregressive parameter $\alpha=0.99999$, and, unless stated otherwise, with unity gain $\lVert \mathbf{w} \rVert^2 = 0$ \SI{}{\deci\bel}. The frequency offsets' random-walk and linear drift parameters as well as the channel time-variance were set as specified separately for any of the results below. Each simulated data point is the average of 16 runs with $10^6$ steady state iterations in each run. Novel signal, channel, and measurement noise realizations were simulated for each run.

\subsection{Steady-State Performance \& Optimal Step Sizes}
\label{ssec:results_steady_state_tracking_performance_and_optimal_step_sizes}

\begin{figure}[htpb!]
    \centering
    \subfloat[random-walk phase noise, sampling jitter, and channel\label{fig:pn_and_sj_and_q_vs_emse}]{\includegraphics[width=.495\textwidth]{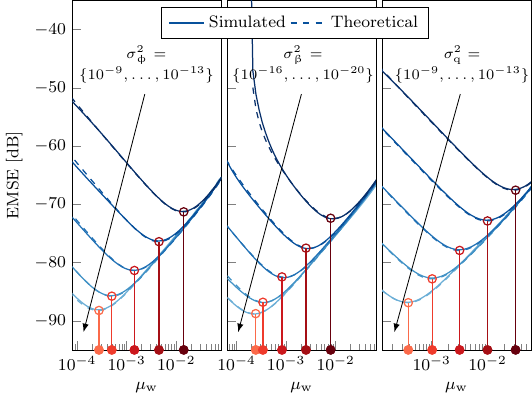}}
    \\
    \subfloat[time-varying carrier frequency offset\label{fig:mu_cfo_vs_emse}]{\includegraphics[width=.495\textwidth]{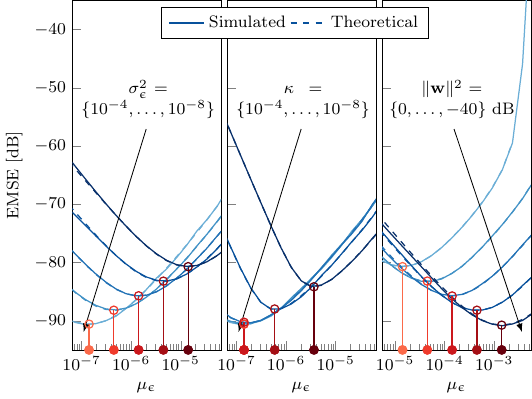}}
    \\
    \subfloat[time-varying sampling frequency offset\label{fig:mu_sfo_vs_emse}]{\includegraphics[width=.495\textwidth]{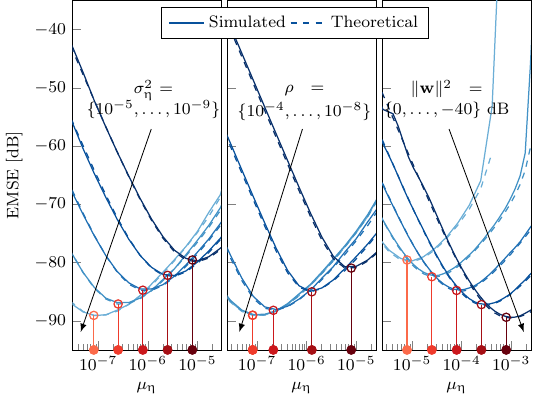}}
    \caption{Theoretical and simulated \glspl{emse} of the \gls{fo-lms} algorithm depending on its step sizes $\mu_\mathrm{w}$,  $\mu_\upepsilon$, and $\mu_\upeta$ for time-varying channels with different rates of $\sigma^2_\mathrm{q}$, $\sigma^2_\upphi$, $\sigma^2_\upbeta$, $\sigma^2_\upepsilon$, $\sigma^2_\upeta$, $\kappa$, $\rho$, and gain $\lVert \mathbf{w} \rVert^2$. Theoretically estimated optimal step sizes for the different cases are plotted on the x-axes.}
    \label{fig:mu_vs_emse}
\end{figure}

Fig.~\ref{fig:mu_vs_emse} compares the theoretical and simulated steady-state \glspl{emse} (i.e., after the algorithm has converged) of the \gls{fo-lms} algorithm over a wide range of step sizes $\mu_\mathrm{w}$, $\mu_\upepsilon$ and $\mu_\upeta$ with the channel and frequency offsets varying at different rates and at different channel gains. Furthermore, given complete knowledge of all of the system parameters, the estimated optimal step sizes $\mu^\mathrm{opt}_\mathrm{w}$, $\mu^\mathrm{opt}_\upepsilon$, and $\mu^\mathrm{opt}_\upeta$ are plotted on the x-axes of Fig.~\ref{fig:pn_and_sj_and_q_vs_emse},~\ref{fig:mu_cfo_vs_emse}, and \ref{fig:mu_sfo_vs_emse}, respectively.

The set of results illustrated in Fig.~\ref{fig:mu_vs_emse} analyze the impact of time-varying channel parameters separately. That is, only one of the parameters $\sigma^2_\upphi$, $\sigma^2_\upbeta$, $\sigma^2_\mathrm{q}$, $\sigma^2_\upepsilon$, $\kappa$, $\sigma^2_\upeta$, and $\rho$ was set to a non-zero value at a time. The results demonstrate that the more rapidly the channel or either frequency offset changes over time, the higher the minimal achievable \gls{emse} becomes. This simply reflects that the performance of the \gls{fo-lms} algorithm is negatively affected by the errors in tracking the time-varying random and cyclic nonstationarities. The right-hand side results in Fig.~\ref{fig:mu_cfo_vs_emse} and \ref{fig:mu_sfo_vs_emse} also demonstrate how the channel gain affects the \glspl{emse} related to compensating the frequency offset drifts.

From these results, it is clear that in order to reach the minimal achievable \gls{emse}, the step sizes need to be carefully selected as the optimal step sizes in Fig.~\ref{fig:mu_vs_emse} span over four orders of magnitude. To that end, it is also an important observation that the estimated \glspl{emse} are accurate, which in turn means that the optimal step size estimation equations in~\eqref{eq:optimal_step_sizes} are appropriate and the estimated step sizes $\mu^\mathrm{opt}_\mathrm{w}$, $\mu^\mathrm{opt}_\upepsilon$, and $\mu^\mathrm{opt}_\upeta$, which are illustrated on the x-axes of Fig.~\ref{fig:mu_vs_emse}, indeed result in minimum achievable \glspl{emse}.

\begin{figure}[htpb!]
    \centering
    \subfloat[
    time-varying channel and carrier frequency offset
    \label{fig:cfo_lms_grid}]{\includegraphics[width=.4825\textwidth]{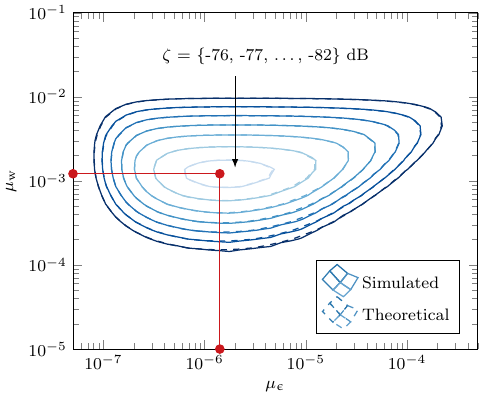}}
    \\
    \subfloat[
    time-varying channel and sampling frequency offset
    \label{fig:sfo_lms_grid}]{\includegraphics[width=.4825\textwidth]{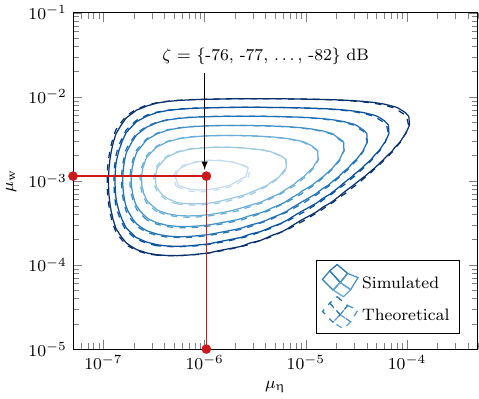}}
    \caption{
    Coupling effect of the \gls{fo-lms} algorithm update equations. Theoretically estimated optimal step sizes are plotted on the corresponding axes. The minimum achievable \gls{emse} is \SI{-82.5}{\decibel}.
    }
    \label{fig:cfo_sfo_grids}
\end{figure}

The results in Fig.~\ref{fig:cfo_sfo_grids} illustrate the coupling effect between the update equations. In both Fig.~\ref{fig:cfo_lms_grid} and Fig. \ref{fig:sfo_lms_grid}, the channel perturbations' variance was set to $\sigma^2_\mathrm{q}=10^{-12}$. In Fig.~\ref{fig:cfo_lms_grid}, the frequency offset drifts were set to $\sigma^2_\upphi = 10^{-12}$, $\sigma^2_\upepsilon = 10^{-6}$, $\kappa=10^{-5}$, $\sigma^2_\upbeta = 0$, $\sigma^2_\upeta = 0$, and $\rho = 0$ with $\mu_\upeta = 0$. Oppositely in Fig.~\ref{fig:sfo_lms_grid}, the frequency offset drifts were set to $\sigma^2_\upphi = 0$, $\sigma^2_\upepsilon = 0$, $\kappa=0$, $\sigma^2_\upbeta = 10^{-19}$, $\sigma^2_\upeta = 10^{-7}$, and $\rho = 5\cdot10^{-6}$ with $\mu_\upepsilon = 0$. The diagonal \gls{emse} curves in the lower right-hand side corners of Fig.~\ref{fig:cfo_lms_grid} and~\ref{fig:sfo_lms_grid} are of especial interest, since for these step size configurations the \gls{emse} is dominated by the coupling effects. These results illustrate how the channel update step size $\mu_\mathrm{w}$ needs to increase when either of the frequency offset update step sizes, $\mu_\upepsilon$ or $\mu_\upeta$, increases. The results also indicate that the expressions capture these effects with good accuracy in both estimating the \gls{emse} and calculating the optimal step sizes.

\begin{figure}[t!]
	\centerline{\includegraphics[width=0.495\textwidth]{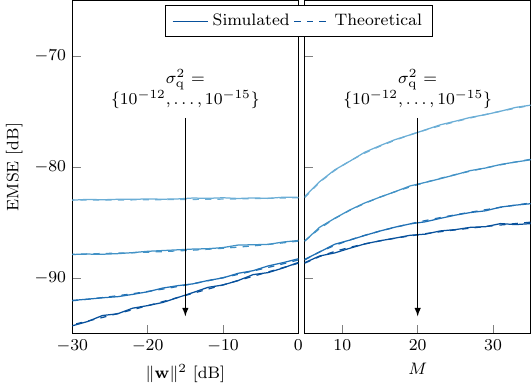}}
	\caption{\Gls{emse} of the \gls{fo-lms} algorithm at optimal step sizes depending on the number of taps $M$ and channel gain $\lVert\mathbf{w}\rVert^2$. The channel and frequency offsets are all time-varying with $\sigma^2_\mathrm{q}$ varied in the range $ \{10^{-12}, \dots, 10^{-15}\}$.}
	\label{fig:M_and_G_vs_emse}
\end{figure}

Finally, we analyze the \gls{fo-lms}'s performance when the channel and frequency offsets all change in time. To that end, the carrier and sampling frequency drifts were set to the frequency offset drifts were set to $\sigma^2_\upphi = 10^{-14}$, $\sigma^2_\upepsilon = 10^{-8}$, $\kappa=10^{-8}$, $\sigma^2_\upbeta = 10^{-21}$, $\sigma^2_\upeta = 10^{-10}$, and $\rho = 10^{-8}$, while the channel time-variance was analyzed in the range $\sigma^2_\mathrm{q}= \{10^{-12}, \dots, 10^{-15}\}$. The channel gain $\lVert\mathbf{w}\rVert^2$ and number of filter taps $M$ in the \gls{fo-lms} algorithm were also varied. Here $M$ denotes only the number of filter taps used by the \gls{fo-lms} to model the channel, not the number of taps actually used to simulate it. The channel was still simulated with 5 taps, as in the rest of the paper. The \glspl{emse} were calculated and simulated using the estimated optimal step sizes.

The results are plotted in Fig.~\ref{fig:M_and_G_vs_emse} and show how the channel gain and number of filter taps used in \gls{fo-lms} impact the minimal achievable \gls{emse}. In alignment with the results in Fig.~\ref{fig:mu_cfo_vs_emse} and \ref{fig:mu_sfo_vs_emse}, the weaker the received signal is that the algorithm needs to match the known signal with, the lower the minimum achievable \gls{emse} becomes. On the opposite, the more filter taps the algorithm uses to match the known signal to that received, the higher the minimum achievable \gls{emse} becomes. These findings hold for adaptive filters in general and various methods have been proposed for reducing impact of the latter when using classical adaptive filters within sparse systems~\cite{chen2009sparse, gu2009l_}. Corresponding research would also be highly fitting to improve the \gls{fo-lms}'s performance and reduce its computational complexity in such cases. However, extending the \gls{fo-lms} algorithm to optimize performance in sparse systems is out of the scope of this paper.

Altogether the results in Figs.~\ref{fig:mu_vs_emse}, \ref{fig:cfo_sfo_grids}, and \ref{fig:M_and_G_vs_emse} exemplify the different scenarios that can arise in practice and give insights into the \gls{fo-lms}'s capability in estimating those. The results also highlight the significance of the step sizes on the algorithm's performance. Given that in practice the system parameters are likely not known and cannot be used to calculate the optimal step sizes, a method to adjust the step sizes at run-time is necessary.

\subsection{Variable Step Sizes}
\label{ssec:results_variable_step_sizes}

In this subsection, we compare the steady state performances of the \gls{fo-lms} and \gls{vss-fo-lms} algorithms in time-varying environments. The results are obtained by running the \gls{fo-lms} algorithm with optimal step sizes calculated based on~\eqref{eq:optimal_step_sizes} and by letting the \gls{vss-fo-lms} algorithm self-adjust its step sizes based on~\eqref{eq:mu_cir_adjusted} and \eqref{eq:variable_fo_step_sizes}. The \gls{vss-fo-lms} algorithm is run in two configurations: with perfect knowledge of the measurement noise variance and with the measurement noise variance estimated during run-time based on~\eqref{eq:noise_variance_estimate}. Theoretically estimated \glspl{emse} of \gls{fo-lms} at optimal step sizes are also plotted for verification. Unless stated otherwise, the \gls{vss-fo-lms} algorithm was initialized with its step sizes limited to $\mu^\mathrm{max}_\mathrm{w} = 10^{-1}$, $\mu^\mathrm{min}_\mathrm{w} = 10^{-5}$, $\mu^\mathrm{max}_\upepsilon = 10^{-3}$, $\mu^\mathrm{min}_\upepsilon = 10^{-9}$, and $\mu^\mathrm{max}_\upeta = 10^{-3}$, $\mu^\mathrm{min}_\upeta = 10^{-9}$, and its forgetting factors set to $\lambda_\mathrm{e}=0.9999$, $\lambda_\mathrm{y}=0.99$, $\lambda_\upepsilon=0.9999$, $\lambda_\upeta=0.9999$, and $\lambda_\mathrm{R}=0.99$. Also, and unless specified otherwise, the channel variance was set to $\sigma^2_\mathrm{q}=10^{-13}$, the carrier frequency offset parameters to $\sigma^2_\upphi = 10^{-13}$, $\sigma^2_\upepsilon = 10^{-8}$, and $\kappa=10^{-8}$, and the sampling frequency offset parameters to $\sigma^2_\upbeta = 10^{-20}$, $\sigma^2_\upeta = 10^{-9}$, and $\rho=10^{-8}$.

\begin{figure}[htpb!]
    \centering
    \subfloat[time-varying channel\label{fig:vss_Pq_vs_emse}]{\includegraphics[width=.495\textwidth]{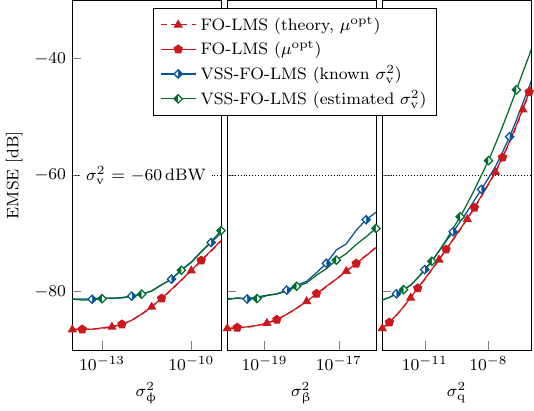}}
    \\
    \subfloat[random-walk frequency offsets\label{fig:DFcfo_and_DFsfo_vs_emse}]{\includegraphics[width=.495\textwidth]{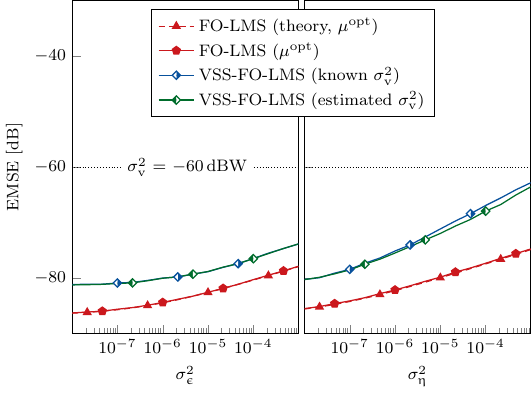}}
    \\
    \subfloat[linear drift frequency offsets\label{fig:vss_DFcfo_DFsfo_vs_emse}]{\includegraphics[width=.495\textwidth]{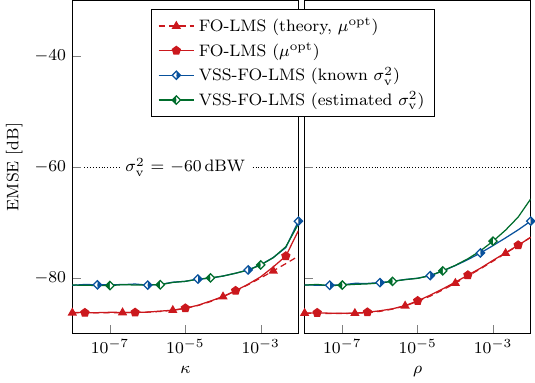}}
    \caption{\Glspl{emse} of the \gls{fo-lms} and \gls{vss-fo-lms} algorithms when estimating time-varying channels depending on the rate of change in the channel and frequency offsets.
    }
    \label{fig:vssfolms_vs_folms}
\end{figure}

The results in Fig.~\ref{fig:vssfolms_vs_folms} show how the two algorithms perform depending on the channel perturbations' variance and depending on the different time-varying components of the frequency offsets. In all cases, it is evident that the \gls{fo-lms} algorithm configured with optimal step sizes outperforms the variable step size solutions, as it achieves lower \gls{emse} regardless of the system configuration. Furthermore, the \gls{vss-fo-lms} algorithm performs better when it perfectly knows the measurement noise variance compared to when it estimates that variance, except when the noise floor is increased due to sampling jitter. Ultimately, the more information the algorithms are provided with about the system beforehand, the smaller the accumulated estimation error is. However, considering the lack of \textit{a priori} knowledge that the proposed \gls{vss-fo-lms} algorithm has about the system and that it reduces the \gls{emse} to below the measurement noise floor, which is at \SI{-60}{\deci\belwatt}, it performs comparatively well.

Finally, as demonstrated earlier in Fig.~\ref{fig:mu_vs_emse}, the optimal \gls{fo-lms} step sizes change several orders of magnitude within the studied system parameter ranges. Yet, as demonstrated by the results in Fig.~\ref{fig:vssfolms_vs_folms}, the \gls{vss-fo-lms} algorithm can be configured with a single set of forgetting factors to reach a good level of performance despite the extent of change in the system parameters. This indicates that the selection of the forgetting factors is not as sensitive to the system parameters as is the selection of step sizes. Which ultimately is the aim of incorporating the variable step sizes adjustment method to the \gls{fo-lms} algorithm, so that the algorithm can be straightforwardly used in practical scenarios where the system parameters are not known in advance and/or change quickly.

\begin{figure}[htpb]
	\centerline{\includegraphics[width=0.495\textwidth]{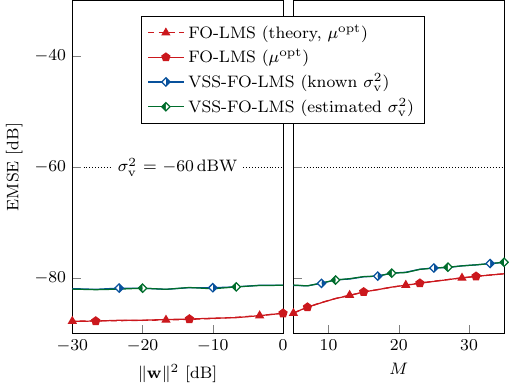}}
	\caption{\Glspl{emse} when estimating time-varying channels with different path losses and when using different filter lenghts.}
	\label{fig:vss_M_and_G_vs_emse}
\end{figure}

\begin{figure}[htpb]
	\centerline{\includegraphics[width=0.495\textwidth]{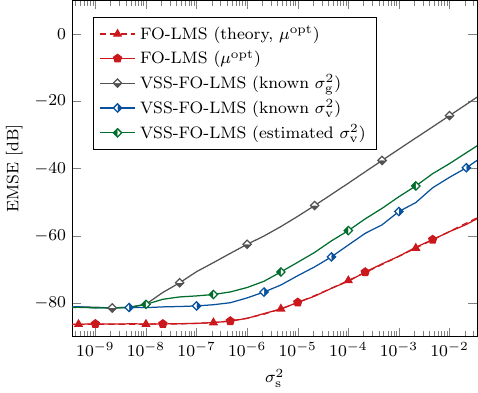}}
	\caption{\Glspl{emse}
    of the \gls{fo-lms} and \gls{vss-fo-lms} algorithms
    when the measurement noise $v(n)$ includes a background signal $s(n)$.}
	\label{fig:vss_Ps_vs_emse}
\end{figure}

Fig.~\ref{fig:vss_M_and_G_vs_emse} demonstrates how the \gls{vss-fo-lms} performs in comparison to the \gls{fo-lms} depending on the path loss and the number of filter taps used to model the channel. The results indicate a consistent performance from the \gls{vss-fo-lms} regardless of the extent of these parameters as, again, the \gls{emse} is reduced to well below the noise floor.

So far we have assumed the background signal $s(n)$ in \eqref{eq:d} to be negligible so that the measurement noise is equal to the receiver noise floor with $\sigma^2_\mathrm{v} = \sigma^2_\mathrm{g}$. The results in Fig.~\ref{fig:vss_Ps_vs_emse} demonstrate how the \gls{vss-fo-lms} algorithm performs when that assumption changes with the introduction of a white Gaussian background signal so that $\sigma^2_\mathrm{v} = \sigma^2_\mathrm{g} + \sigma^2_\mathrm{s}$. Then, as the received background signal power level increases, \gls{vss-fo-lms} performs best when the exact combined power level of the received background signal and receiver noise floor is known. However, if the algorithm relies only on knowledge of the receiver noise floor level, its performance in the corresponding cases deteriorates. If the \gls{vss-fo-lms} algorithm relies on the measurement noise level estimation based on~\eqref{eq:noise_variance_estimate}, it outperforms the algorithm with the misleading noise floor estimate and slightly loses out to that with perfect knowledge of the total measurement noise level.

\begin{figure}[t!]
	\centerline{\includegraphics[width=0.495\textwidth]{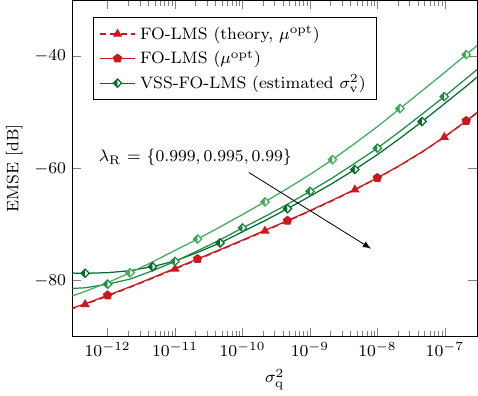}}
	\caption{\Gls{emse} of the \gls{vss-fo-lms} algorithm depending on the forgetting factor $\lambda_\mathrm{R}$ and channel time-variance $\sigma^2_\mathrm{q}$.}
	\label{fig:vss_lambdaR_vs_emse}
\end{figure}

\begin{figure}[tpb]
	\centerline{\includegraphics[width=0.495\textwidth]{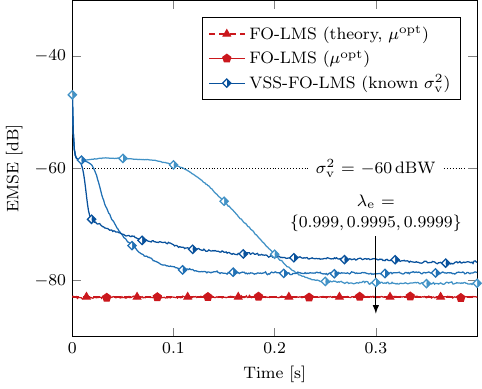}}
	\caption{Convergence of the \gls{vss-fo-lms} algorithm depending on the forgetting factor $\lambda_\mathrm{e}$. Steady-state performance of \Gls{fo-lms} with optimal step sizes is shown to illustrate the minimal achievable \gls{emse} only.}
	\label{fig:vss_time_vs_emse}
\end{figure}

Finally, while the results in Figs.~\ref{fig:vssfolms_vs_folms}, \ref{fig:vss_M_and_G_vs_emse}, and \ref{fig:vss_Ps_vs_emse} showed that the \gls{vss-fo-lms} algorithm performs in a stable manner for all of the studied cases with just a single set of forgetting factors, the forgetting factors do affect the algorithm's performance and at least some care needs to be put into selecting their values in practice. For example, $\lambda_\mathrm{R}$, which controls the exponential decay of the cross-correlation vector estimate in~\eqref{eq:cross_correlation_vector_estimation}, introduces a moderate trade-off to the algorithm's steady-state misadjustment and tracking performance as illustrated in Fig~\ref{fig:vss_lambdaR_vs_emse}. Similarly, $\lambda_\mathrm{e}$, which controls the exponential decay of the estimation error estimate in, presents a trade-off between steady-state \gls{emse} and convergence rate as illustrated in Fig.~\ref{fig:vss_time_vs_emse}.

\section{Conclusion}
\label{sec:conclusion}

\glsresetall
In this paper, we studied the steady-state performance of the \gls{fo-lms} algorithm in the presence of time-varying random and cyclic nonstationarities. In particular, we provided expressions for estimating the algorithm's \gls{emse} under time varying channels as well as time varying carrier and sampling frequency offsets. We also provided expressions for calculating the algorithm's optimal step sizes that minimize the \gls{emse} but the evaluation of which requires the system parameters to be known. To simplify practical applications where such knowledge cannot be relied on, we proposed a \gls{vss-fo-lms} algorithm that self-adjusts its step sizes based on its performance.

We verified the \gls{emse} and optimal step size expressions of the \gls{fo-lms} algorithm and evaluated the performance of the proposed \gls{vss-fo-lms} algorithm through simulations. The \gls{emse} expressions allow us to accurately estimate the algorithm's performance when the algorithm is in steady state and the optimal step size expressions offer a scalable and low-overhead way of tuning real-time applications when the system parameters are known. The proposed \gls{vss-fo-lms} method offers a straightforward way of letting the algorithm itself find a sensible configuration of step sizes when the system parameters are not know, albeit at a moderate performance penalty compared to operation with optimal step sizes. In a follow-up paper~\cite{parlin2025dfkic}, we analyze how \gls{vss-fo-lms} performs when processing measurement data captured with \glspl{sdr}.

\appendix
\label{ap:emse}

Using \eqref{eq:d}, \eqref{eq:updates}, and \eqref{eq:parameter_differences} we obtain the following recursions for the parameter errors:
\begin{subequations}\label{eq:parameter_differences_recursion}
\begin{align}\label{eq:cir_difference_recursion}
    \tilde{\mathbf{w}}_{n+1} &= \tilde{\mathbf{w}}_{n} - \mu_\mathrm{w} \mathbf{y}_n e^{j \phi(n) } e^{*}(n) + \mathbf{a}_n,\\\label{eq:cfo_difference_recursion}
    \tilde{\epsilon}(n+1) &= \tilde{\epsilon}(n) - \mu_{{\upepsilon}} \mathbf{w}^{H}_{n} \mathbf{y}_n e^{j \phi(n) } e^{*}(n) + b(n),\\\label{eq:sfo_difference_recursion}
    \tilde{\eta}(n+1) &= \tilde{\eta}(n) - \mu_{{\upeta}} \mathbf{w}^{H}_{n} \mathbf{y}'_n  e^{j \phi(n) }e^{*}(n) + c(n).
\end{align}
\end{subequations}
Above, considering that for small time offsets the sampling jitter can be modeled as phase noise~\cite{syrjala2010sampling}, relying on linear approximation of both the phase noise and sampling jitter, and considering that $\alpha \approx 1$, we define
\begin{multline}
    \mathbf{a}_n \delequal \mathbf{w}^\mathrm{o}\omega_\upphi(n)
    + \mathbf{w}^\mathrm{o}\sfrac{\omega_\upbeta(n)}{T_\mathrm{s}} \\
    + \mathbf{w}^\mathrm{o} \tilde{\epsilon}(n) + \mathbf{w}^\mathrm{o} \tilde{\eta}(n)
    + \mathbf{q}_n
\end{multline}
with $b(n) \delequal \omega_\upepsilon(n) + \kappa$ and $c(n) \delequal \omega_\upeta(n) + \rho$. Then, similarly to the \textit{a priori} errors in \eqref{eq:a_priori_errors}, we define the \textit{a posteriori} errors%
\begin{subequations}\label{eq:a_posteriori_errors_0}
\begin{align}
    e^\mathrm{p}_\mathrm{w}(n) &= (\tilde{\mathbf{w}}_{n+1} - \mathbf{a}_n)^{H} \mathbf{y}_n e^{j\phi(n)},\\
    e^\mathrm{p}_\mathrm{\upepsilon}(n) &= \mathbf{w}^{H}_{n} \mathbf{y}_n \left( \tilde{\epsilon}(n+1) - b(n) \right) e^{j\phi(n)},\\
    e^\mathrm{p}_\mathrm{\upeta}(n) &= \mathbf{w}^{H}_{n} \mathbf{y}'_n \left( \tilde{\eta}(n+1) - c(n) \right) e^{j\phi(n)}.
\end{align}
\end{subequations}
Shifting the terms $\mathbf{a}_n$, $b(n)$, and $c(n)$ in
(\ref{eq:parameter_differences_recursion}) to the left-hand side, taking the conjugate transpose, and multiplying the resulting equations from the right by, respectively, $\mathbf{y}_n e^{j \phi(n) }$, $\mathbf{w}^{H}_{n} \mathbf{y}_n e^{j\phi(n)}$, and $\mathbf{w}^{H}_{n} \mathbf{y}'_n e^{j\phi(n)}$ we also find that
\begin{subequations}\label{eq:a_posteriori_errors_1}
\begin{align}
    e^\mathrm{p}_\mathrm{w}(n) =\ &e^\mathrm{a}_\mathrm{w}(n) - \mu_\mathrm{w} \lVert \mathbf{y}_n \rVert^2 e(n) ,\\
    e^\mathrm{p}_\mathrm{\upepsilon}(n) =\ &e^\mathrm{a}_\mathrm{\upepsilon}(n) - \mu_\upepsilon \lvert \mathbf{w}^{H}_{n} \mathbf{y}_n  \rvert^2 e^{*}(n),\\
    e^\mathrm{p}_\mathrm{\upeta}(n) =\ &e^\mathrm{a}_\mathrm{\upeta}(n) - \mu_\upepsilon \lvert \mathbf{w}^{H}_{n} \mathbf{y}'_n \rvert^2 e^{*}(n).
\end{align}
\end{subequations}
The instantaneous total error can therefore be written as
\begin{subequations}
\begin{align}
    e(n) =\ & \mu^{-1}_\mathrm{w} \tilde{\mu}_\mathrm{w}(n) \left[ e^\mathrm{a}_\mathrm{w}(n) - e^\mathrm{p}_\mathrm{w}(n) \right],\\
    e(n) =\ & \mu^{-1}_\upepsilon \tilde{\mu}_\upepsilon(n) \left[ e^\mathrm{a}_\upepsilon(n) - e^\mathrm{p}_\upepsilon(n) \right]^*,\\
    e(n) =\ & \mu^{-1}_\upeta \tilde{\mu}_\upeta(n) \left[ e^\mathrm{a}_\upeta(n) - e^\mathrm{p}_\upeta(n) \right]^*,
\end{align}
\end{subequations}
where we have defined $\tilde{\mu}_\mathrm{w}(n) = \left(\lVert \mathbf{y}_n \rVert^2\right)^+$, $\tilde{\mu}_\upepsilon(n) = \left(\lvert \mathbf{w}^{H}_{n} \mathbf{y}_n  \rvert^2\right)^+$, and $\tilde{\mu}_\upeta(n) = \left(\lvert \mathbf{w}^{H}_{n} \mathbf{y}'_n \rvert^2\right)^+$ in terms of the scalar pseudoinverses. Substituting the right-hand sides of the above into \eqref{eq:parameter_differences_recursion} and by evaluating the energies on both sides of the resulting equations we find that
\begin{subequations}\label{eq:parameter_differences_recursion_2}
\begin{align}\label{eq:cir_energy_relation}
    |\tilde{\mathbf{w}}_{n+1} - \mathbf{a}_n|^2 &+ \tilde{\mu}_\mathrm{w}(n)|e^\mathrm{a}_\mathrm{w}(n)|^2 =\\\nonumber &|\tilde{\mathbf{w}}_{n}|^2 + \tilde{\mu}_\mathrm{w}(n)|e^\mathrm{p}_\mathrm{w}(n)|^2,\\\label{eq:cfo_energy_relation}
    |\tilde{\epsilon}(n+1) - b(n)|^2 &+ \tilde{\mu}_\upepsilon(n)|e^\mathrm{a}_\mathrm{\upepsilon}(n)|^2 =\\\nonumber &|\tilde{\epsilon}(n)|^2 + \tilde{\mu}_\upepsilon(n)|e^\mathrm{p}_\mathrm{\upepsilon}(n)|^2,\\\label{eq:sfo_energy_relation}
    |\tilde{\eta}(n+1) - c(n)|^2 &+ \tilde{\mu}_\upeta(n)|e^\mathrm{a}_\mathrm{\upeta}(n)|^2 =\\\nonumber &|\tilde{\eta}(n)|^2 + \tilde{\mu}_\upeta(n)|e^\mathrm{p}_\mathrm{\upeta}(n)|^2.
\end{align}
\end{subequations}

Considering that $\mathbf{q}_n$, $\tilde{\epsilon}(n)$, $\tilde{\eta}(n)$, $\omega_\upphi(n)$, and $\omega_\upbeta(n)$ are independent, that in steady state $E\left[\lvert\tilde{\mathbf{w}}_{n+1}\rvert^2\right] = E\left[\lvert\tilde{\mathbf{w}}_{n}\rvert^2\right]$, and relying on the self-induced nonstationarity tracking~\cite{parlin2023estimating} and nonstationary channel tracking concepts~\cite{yousef2001unified}, the steady state expected relation of \eqref{eq:cir_energy_relation} can be expressed as
\begin{subequations}\label{eq:emse_relation_step_1}
\begin{multline}\label{eq:cir_relation_step_1}
    E\left[ \tilde{\mu}_\mathrm{w}(n)|e^\mathrm{a}_\mathrm{w}(n)|^2 \right] =\
    \mathrm{Tr}(\mathbf{Q})\\
    + \lVert\mathbf{w}^\mathrm{o}\rVert^2 E\left[ \lvert \omega_\upphi(n) \rvert^2 \right]
    + \lVert\mathbf{w}^\mathrm{o}\rVert^2 E\left[ \lvert \sfrac{\omega_\upbeta(n)}{ T^{}_\mathrm{s} }\rvert^2 \right]\\
    + E\left[ \tilde{\mu}_\mathrm{w}(n)\lvert e^\mathrm{a}_\upepsilon(n) \rvert^2 \right]\
    + E\left[ \tilde{\mu}_\mathrm{w}(n)\lvert e^\mathrm{a}_\upeta(n) \rvert^2 \right]\\
    + E\left[ \tilde{\mu}_\mathrm{w}(n)\lvert e^\mathrm{a}_\mathrm{w}(n) - \mu_\mathrm{w} \tilde{\mu}^{-1}_\mathrm{w}(n) e(n) \rvert^2 \right].
\end{multline}
Similarly, assuming that in steady state $E\left[|\tilde{\epsilon}(n+1)|^2\right] = E\left[|\tilde{\epsilon}(n)|^2\right]$ and $E\left[|\tilde{\eta}(n+1)|^2\right] = E\left[|\tilde{\eta}(n)|^2\right]$, the steady-state relations of \eqref{eq:cfo_energy_relation} and \eqref{eq:sfo_energy_relation} can be expressed as
\begin{multline}\label{eq:cfo_relation_step_1}%\addtocounter{equation}{1}
    E\left[\tilde{\mu}_\upepsilon(n)|e^\mathrm{a}_\mathrm{\upepsilon}(n)|^2\right] \approx \\
    2 T_\mathrm{s}^2 E\left[\left| \omega_\upepsilon(n) \right|^{2}\right]
    + 2 \mu_\mathrm{w} \mu_{\upepsilon}^{-1} \kappa^2 T_\mathrm{s}^{2} E\left[ \tilde{\mu}_\upepsilon(n) \tilde{\mu}^{-1}_\mathrm{w}(n) \right]\\
    + E\left[\tilde{\mu}_\upepsilon(n)|e^\mathrm{a}_\mathrm{\upepsilon}(n) - \mu_\upepsilon \tilde{\mu}^{-1}_\upepsilon(n)  e^{*}(n)|^2\right],
\end{multline}
\begin{multline}\label{eq:sfo_relation_step_1}
    E\left[\tilde{\mu}_\upeta(n)|e^\mathrm{a}_\mathrm{\upeta}(n)|^2\right] \approx \\
    2 T_\mathrm{s}^2 E\left[\left| \omega_\upeta(n) \right|^{2}\right]
    + 2 \mu_\mathrm{w} \mu_{\upeta}^{-1} \rho^2 T_\mathrm{s}^{2} E\left[ \tilde{\mu}_\upeta(n) \tilde{\mu}^{-1}_\mathrm{w}(n) \right]\\
    + E\left[\tilde{\mu}_\upeta(n)|e^\mathrm{a}_\mathrm{\upeta}(n) - \mu_\upeta \tilde{\mu}^{-1}_\upeta(n) e^{*}(n)|^2\right]\\
    + \frac{2 \mu^2_\mathrm{w}}{T^{2}_\mathrm{s}} E\left[ \frac{ \lvert \omega_\upbeta(n)\rvert^2}{\tilde{\mu}^{}_\upeta(n)} \right]
    + \frac{2 \mu^2_\mathrm{w} T^2_\mathrm{s} }{ \mu^{}_\upeta } E\left[ \frac{\tilde{\mu}_\upeta(n) \left| \omega_\upeta \right|^2}{ \tilde{\mu}^2_\mathrm{w}(n) } \right],
\end{multline}
\end{subequations}
where the last two terms in \eqref{eq:sfo_relation_step_1} effectively account for the reduced \gls{snr} caused by the clock jitter and drift~\cite{da2002jitter, lohning2007effects, towfic2010sampling}. Rigorous derivation of \eqref{eq:emse_relation_step_1} is omitted herein but the simulations in Section~\ref{sec:results} demonstrate their applicability.

Expanding the last terms on the right-hand sides of \eqref{eq:cir_relation_step_1}, \eqref{eq:cfo_relation_step_1}, and \eqref{eq:sfo_relation_step_1}, rearranging the equations, and dividing by the respective step sizes, we get
\begin{subequations}
\begin{multline}
    2 E\left[ e(n) e^\mathrm{a}_\mathrm{w}(n) \right] =
    \mu_\mathrm{w} E\left[ \frac{ \lvert e(n) \rvert^2 }{ \tilde{\mu}^{}_\mathrm{w}(n) }\right]
    + \frac{
    \mathrm{Tr}(\mathbf{Q})
    }{\mu_\mathrm{w}}\\
    + \frac{
    \lVert\mathbf{w}^\mathrm{o}\rVert^2 E\left[ \lvert \omega_\upphi(n) \rvert^2 \right]
    }{\mu_\mathrm{w}}
    + \frac{
    \lVert\mathbf{w}^\mathrm{o}\rVert^2 E\left[ \lvert \omega_\upbeta(n)\rvert^2 \right]
    }{\mu_\mathrm{w} T^2_\mathrm{s}} \\
    + \frac{ E\left[ \tilde{\mu}_\mathrm{w}(n)|e^\mathrm{a}_{\upepsilon}(n)|^2 \right] }{\mu^{}_w}
    + \frac{ E\left[ \tilde{\mu}_\mathrm{w}(n)|e^\mathrm{a}_{\upeta}(n)|^2 \right] }{ \mu^{}_w },
\end{multline}
\begin{multline}
    2 E\left[ e^*(n) e^\mathrm{a}_\mathrm{\upepsilon}(n) \right] =
    \mu_\upepsilon E\left[ \frac{ |e(n)|^2 }{ \tilde{\mu}^{}_\upepsilon(n) } \right] \\
    + \frac{2 T_\mathrm{s}^2 E\left[\left| \omega_\upepsilon(n) \right|^{2}\right]}{\mu_{\upepsilon}}
    + \frac{2 \mu_\mathrm{w} \kappa^{2} T^2_\mathrm{s}}{ \mu_{\upepsilon}^{2}} E\left[\frac{\tilde{\mu}_\upepsilon(n)}{\tilde{\mu}_\mathrm{w}(n)} \right],
\end{multline}
\begin{multline}
    2 E\left[ e^*(n) e^\mathrm{a}_\mathrm{\upeta}(n) \right] =
    \mu_\upeta E\left[ \frac{ |e(n)|^2 }{ \tilde{\mu}_\upeta(n) } \right]
    + \frac{2 T_\mathrm{s}^2 E\left[\left| \omega_\upeta(n) \right|^{2}\right]}{\mu_{\upeta}} \\
    + \frac{2 \mu_\mathrm{w} \rho^2 T^2_\mathrm{s}}{ \mu_{\upeta}^{2}} E\left[\frac{\tilde{\mu}_\upeta(n)}{\tilde{\mu}_\mathrm{w}(n)} \right]
    + \frac{ 2 \mu^2_\mathrm{w} }{ \mu_{\upeta} T^2_\mathrm{s} }E\left[\frac{\left|{ \omega_\upbeta(n) }\right|^{2}}{\tilde{\mu}_\upeta(n)}\right]\\
    + \frac{2 \mu^2_\mathrm{w} T^2_\mathrm{s} }{ \mu^{2}_\upeta } E\left[ \frac{\tilde{\mu}_\upeta(n) \left| \omega_\upeta \right|^2}{ \tilde{\mu}^2_\mathrm{w}(n) } \right].
\end{multline}
\end{subequations}
Substituting \eqref{eq:e_total_a_priori} into the above while also assuming that the noise sequence $v(n)$ is stationary and statistically independent of the \textit{a priori} errors $e^\mathrm{a}_\mathrm{w}(n)$, $e^\mathrm{a}_{\upepsilon}(n)$, and $e^\mathrm{a}_{\upeta}(n)$, leads to \eqref{eq:e_a_relations} at the top of the next page. To proceed, we make the following assumptions that are commonly used in adaptive filter performance analysis~\cite{jiang2015block, zhang2020steady}.

\begin{assump}\label{assump:steady_state_channel_approaches_true_channel}
In steady state, as $n \to \infty$, the channel estimate is close to the actual channel $\mathbf{w}_{n} \to \mathbf{w}^\mathrm{o}_n$.
\end{assump}
\begin{assump}\label{assump:parameter_error_independence}
In steady state, as $n \to \infty$, the parameter errors $\tilde{\mathbf{w}}_n$, $\tilde{\epsilon}(n)$, and $\tilde{\eta}(n)$ are all statistically independent of $\mathbf{y}_n$, $\mathbf{w}^H_n\mathbf{y}_n$, and $\mathbf{w}^H_n \mathbf{y}'_n$.
\end{assump}

\begin{figure*}[t]
\begin{subequations}\label{eq:e_a_relations}
\begin{multline}\label{eq:e_a_relations_cir}
    2 E\left[\left|{e^a_\mathrm{w}(n)}\right|^{2}\right] =
    \frac{
    \mathrm{Tr}(\mathbf{Q})
    }{\mu_\mathrm{w}}
    + \frac{
    \lVert\mathbf{w}^\mathrm{o}\rVert^2 E\left[ \lvert \omega_\upphi(n) \rvert^2 \right]
    }{\mu_\mathrm{w}}
    + \frac{
    \lVert\mathbf{w}^\mathrm{o}\rVert^2 E\left[ \lvert \omega_\upbeta(n)\rvert^2 \right]
    }{\mu_\mathrm{w} T^2_\mathrm{s}}
    + \frac{E\left[\tilde{\mu}_\mathrm{w}(n) \left|e^a_{\upepsilon}(n)\right|^{2}\right]}{\mu_\mathrm{w}}
    + \frac{E\left[\tilde{\mu}_\mathrm{w}(n) \left|e^a_{\upeta}(n)\right|^{2}\right]}{\mu_\mathrm{w}}\\
    + \mu_\mathrm{w}E\left[\frac{\left|{e^a_\mathrm{w}(n)}\right|^{2}}{\tilde{\mu}_\mathrm{w}(n)}\right] 
    + \mu_\mathrm{w}E\left[\frac{\left|{e^a_\mathrm{\upepsilon}(n)}\right|^{2}}{\tilde{\mu}_\mathrm{w}(n)}\right]
    + \mu_\mathrm{w}E\left[\frac{\left|{e^a_\mathrm{\upeta}(n)}\right|^{2}}{\tilde{\mu}_\mathrm{w}(n)}\right] 
    + \mu_\mathrm{w}E\left[\frac{\left|{v(n)}\right|^{2}}{\tilde{\mu}_\mathrm{w}(n)}\right] 
\end{multline}
\begin{multline}\label{eq:e_a_relations_cfo}
    2 E\left[\left|{e^a_\mathrm{\upepsilon}(n)}\right|^{2}\right] =
    \frac{2 T^2_\mathrm{s} E\left[\left| \omega_\upepsilon(n) \right|^{2}\right]}{\mu_{\upepsilon}}
    + \frac{2 \mu_\mathrm{w} \kappa^{2} T^2_\mathrm{s}}{ \mu_{\upepsilon}^{2}} E\left[\frac{\tilde{\mu}_\upepsilon(n)}{\tilde{\mu}_\mathrm{w}(n)} \right]\\
    + \mu_{\upepsilon}E\left[\frac{\left|{e^a_\mathrm{w}(n)}\right|^{2}}{\tilde{\mu}_\upepsilon(n)}\right]
    + \mu_{\upepsilon}E\left[\frac{\left|{e^a_\mathrm{\upepsilon}(n)}\right|^{2}}{\tilde{\mu}_\upepsilon(n)}\right]
    + \mu_{\upepsilon}E\left[\frac{\left|{e^a_\mathrm{\upeta}(n)}\right|^{2}}{\tilde{\mu}_\upepsilon(n)}\right] 
    + \mu_{\upepsilon}E\left[\frac{\left|{v(n)}\right|^{2}}{\tilde{\mu}_\upepsilon(n)}\right]
\end{multline}
\begin{multline}\label{eq:e_a_relations_sfo}
    2 E\left[\left|{e^a_\mathrm{\upeta}(n)}\right|^{2}\right] =
    \frac{2 T^2_\mathrm{s} E\left[\left| \omega_\upeta(n) \right|^{2}\right]}{\mu_{\upeta}}
    + \frac{2 \mu_\mathrm{w} \rho^2 T^2_\mathrm{s}}{ \mu_{\upeta}^{2}} E\left[\frac{\tilde{\mu}_\upeta(n)}{\tilde{\mu}_\mathrm{w}(n)} \right]
    + \frac{ 2 \mu^2_\mathrm{w} }{ \mu_{\upeta} T^2_\mathrm{s} }E\left[\frac{\left|{\omega_\upbeta(n) }\right|^{2}}{\tilde{\mu}_\upeta(n)}\right]
    + \frac{2 \mu^2_\mathrm{w} T^2_\mathrm{s} }{ \mu^{2}_\upeta } E\left[ \frac{\tilde{\mu}_\upeta(n) \left| \omega_\upeta \right|^2}{ \tilde{\mu}^2_\mathrm{w}(n) } \right] \\
    + \mu_{\upeta}E\left[\frac{\left|{e^a_\mathrm{w}(n)}\right|^{2}}{\tilde{\mu}_\upeta(n)}\right]
    + \mu_{\upeta}E\left[\frac{\left|{e^a_\mathrm{\upepsilon}(n)}\right|^{2}}{\tilde{\mu}_\upeta(n)}\right]
    + \mu_{\upeta}E\left[\frac{\left|{e^a_\mathrm{\upeta}(n)}\right|^{2}}{\tilde{\mu}_\upeta(n)}\right]
    + \mu_{\upeta}E\left[\frac{\left|{v(n)}\right|^{2}}{\tilde{\mu}_\upeta(n)}\right]
\end{multline}
\end{subequations}
\end{figure*}

Then, relying on the assumptions A.\ref{assump:steady_state_channel_approaches_true_channel} and A.\ref{assump:parameter_error_independence} and adhering to the same reasoning that is used for analyzing the steady-state performance of the \gls{lms} adaptive filter~\cite[p. 296]{sayed2003fundamentals}, the following approximations can be made
\begin{subequations}\label{eq:gaussian_white_expansions}
\begin{align}
    E\left[ \frac{ |e^\mathrm{a}_\mathrm{w}(n)|^2 }{ \tilde{\mu}_\mathrm{w}(n) } \right] &\approx (1 + M) \sigma^2_\mathrm{x} \zeta_\mathrm{w},\\
    E\left[ \frac{ |e^\mathrm{a}_\mathrm{w}(n)|^2 }{ \tilde{\mu}_\upepsilon(n) } \right] &\approx \sigma^2_\mathrm{x} \lVert\mathbf{w}^\mathrm{o}_n\rVert^2 \zeta_\mathrm{w},\\
    E\left[ \frac{ |e^\mathrm{a}_\upepsilon(n)|^2 }{ \tilde{\mu}_\upepsilon(n) } \right] &\approx \sigma^2_\mathrm{x} \lVert\mathbf{w}^\mathrm{o}_n\rVert^2 \zeta_\upepsilon,\\
    E\left[ \frac{ |e^\mathrm{a}_\mathrm{w}(n)|^2 }{ \tilde{\mu}_\upeta(n) } \right] &\approx (2 + \sfrac{2}{M}) \sigma^2_\mathrm{x} \lVert\mathbf{w}^\mathrm{o}_n\rVert^2 \zeta_\mathrm{w},\\
    E\left[ \frac{ |e^\mathrm{a}_\upeta(n)|^2 }{ \tilde{\mu}_\upeta(n) } \right] &\approx  \sigma^2_\mathrm{x} \lVert\mathbf{w}^\mathrm{o}_n\rVert^2 \zeta_\upeta.
\end{align}
\end{subequations}

Finally, the system of equations in \eqref{eq:e_a_relations} can now be solved for the individual \glspl{emse} as
\begin{subequations}\label{eq:emse_all_gaussian_complete}
\begin{multline}
    \zeta_\mathrm{w} = \frac{
     \mu_\mathrm{w}M \sigma^{2}_{v} \sigma^{2}_{x}
    + \frac{M \sigma^{2}_{q}}{\mu_\mathrm{w}}
    + \frac{\mu_\upepsilon \lVert\mathbf{w}^\mathrm{o}\rVert^2 \sigma^{2}_{v}}{2 \mu_\mathrm{w}} 
    + \frac{\mu_\upeta \lVert\mathbf{w}^\mathrm{o}\rVert^2 \sigma^{2}_{v}}{\mu_\mathrm{w}}
    }{
    \gamma
    }\\
    + \frac{\frac{ \lVert\mathbf{w}^\mathrm{o}\rVert^2 \sigma^2_\upphi }{\mu_\mathrm{w}}
    + \frac{ \lVert\mathbf{w}^\mathrm{o}\rVert^2 \sigma^2_\upbeta}{\mu_\mathrm{w} T_\mathrm{s}} }{\gamma},
\end{multline}
\begin{equation}
    \zeta_\upepsilon = \frac{
    \mu_{\upepsilon} \sigma^{2}_{x} \lVert\mathbf{w}^\mathrm{o}\rVert^2 \sigma^{2}_{v}
    + \frac{\sigma^2_\upepsilon T_\mathrm{s}^2}{\mu_\mathrm{w} \mu_\upepsilon \sigma^2_\mathrm{x}}
    + \frac{2 \kappa^{2} T_\mathrm{s}^2}{\mu_{\upepsilon}^{2} \sigma^{2}_{x} \lVert\mathbf{w}^\mathrm{o}\rVert^2 }
    }{
    \gamma
    },
\end{equation}
\begin{multline}
    \zeta_\upeta = \frac{
    2 \mu_{\upeta} \sigma^{2}_{x} \lVert\mathbf{w}^\mathrm{o}\rVert^2 \sigma^{2}_{v}
    + \frac{
    \sigma^2_\upeta T_\mathrm{s}^2
    }{\mu_\mathrm{w} \mu_\upeta \sigma^2_\mathrm{x}}
    + \frac{\rho^2 T_\mathrm{s}^2}{(2+\sfrac{2}{M}) \mu_{\upeta}^{2} \sigma^{2}_{x} \lVert\mathbf{w}^\mathrm{o}\rVert^2 }
    }{
    \gamma
    }\\
    + \frac{\frac{ \mu_\mathrm{w} \lVert\mathbf{w}^\mathrm{o}\rVert^2 \sigma^2_\upbeta}{ \mu_\upeta T_\mathrm{s}} +
    \frac{
    \mu_\mathrm{w} \sigma^2_\upeta T_\mathrm{s}^2}{\mu_\upeta^2 \lVert\mathbf{w}^\mathrm{o}\rVert^2}
    }{\gamma},
\end{multline}
\end{subequations}
where
\begin{multline}\label{eq:emse_denominator}
    \gamma = 2
    - \mu_\mathrm{w} \left(1 + M\right) \sigma^2_\mathrm{x}\\
    - \frac{ \mu_\upepsilon }{ \mu_\mathrm{w} } \lVert \mathbf{w}^\mathrm{o} \rVert^2
    - 2\frac{ \mu_\upeta }{ \mu_\mathrm{w} } \left(2 + \frac{2}{M}\right) \lVert \mathbf{w}^\mathrm{o} \rVert^2
\end{multline}
and the negligible cross terms have been discarded. For small step size values, \eqref{eq:emse_all_gaussian_complete} can be accurately approximated as \eqref{eq:emse_all_gaussian}.

% Can use something like this to put references on a page
% by themselves when using endfloat and the captionsoff option.
\ifCLASSOPTIONcaptionsoff
  \newpage
\fi

% references section
\IEEEtriggeratref{37}
\bibliographystyle{IEEEtran}
\bibliography{bibtex/bib/IEEEabrv,bibtex/bib/paper}

% that's all folks
\end{document}